\documentclass[useAMS,usenatbib]{mn2e}

\usepackage{lscape}
\usepackage{epsfig}
\usepackage{times}

\def\gtsim{\mathrel{\spose{\lower.5ex \hbox{$\mathchar"218$}}
     \raise.4ex\hbox{$\mathchar"13E$}}}
\def\ltsim{\mathrel{\spose{\lower.5ex\hbox{$\mathchar"218$}}
     \raise.4ex\hbox{$\mathchar"13C$}}}
\def\aFe{[$\alpha/{\rm Fe}$]~}

\def\Hb{${\rm H}_{\beta}$}

\def\Mgb{{\rm Mg}\,$_b$}
\def\Fe{$\langle {\rm Fe}\rangle$}

\def\ZH{[$Z/{\rm H}$]~}
\def\MgFe{[${\rm MgFe}$]$'$}

\def\Mgd{{\rm Mg}\,$_2$}

\def\Rbd{$r_{\rm{ bd}}$} 

\def\kms{$\rm km\;s^{-1}$}

\def\mas{mag arcsec$^{-2}$}

\def\spose#1{\hbox to 0pt{#1\hss}}

\begin{document}

\title[Bulges in low surface brightness galaxies]{Structure and
  dynamics of galaxies with a low surface-brightness disc -
  II. Stellar populations of bulges\thanks{Based on observations made
    with ESO Telescopes at the La Silla Paranal Observatory under
    programmes 76.B-0375 and 80.B-00754.}}

%\subtitle{}
\author[L. Morelli et al.]{L.~Morelli$^{1,2}$, E.~M.~Corsini$^{1,2}$, 
  A.~Pizzella$^{1,2}$, E. Dalla Bont\`a$^{1,2}$, L.~Coccato$^{3}$,
  \newauthor J.~M\'endez-Abreu$^{4,5}$ and M. Cesetti$^{1,2}$\\
$^1$ Dipartimento di Fisica e Astronomia, Universit\`a di Padova,
  vicolo dell'Osservatorio~3, I-35122 Padova, Italy.\\
$^2$ INAF--Osservatorio Astronomico di Padova,
  vicolo dell'Osservatorio 5, I-35122 Padova, Italy.\\
$^{3}$ European Southern Observatory, Karl-Schwarzschild-Stra$\beta$e 
2, D-85748 Garching bei M\"unchen, Germany.  \\
$^4$ Instituto Astrof\'\i sico de Canarias, 
C/ V\'ia L\'actea s/n, E-38200 La Laguna, Spain\\
$^5$ Departamento de Astrof\'\i sica, Universidad de La Laguna, 
C/ Astrof\'isico Francisco S\'anchez, E-38205 La Laguna, Spain}

%\date{Received..................; accepted...................}
\date{{\it Draft version on \today}}
%\date{Accepted 1988 December 15. Received 1988 December 14; in original form 1988 October 11}
%\pagerange{\pageref{firstpage}--\pageref{lastpage}} \pubyear{2002}

\maketitle

%\label{firstpage}

\begin{abstract}

The radial profiles of the \Hb , Mg, and Fe line-strength indices are
presented for a sample of eight spiral galaxies with a low
surface-brightness stellar disc and a bulge. 
The correlations between the central values of the line-strength
indices and velocity dispersion are consistent to those known for
early-type galaxies and bulges of high surface-brightness galaxies. 
The age, metallicity, and $\alpha/$Fe enhancement of the stellar
populations in the bulge-dominated region are obtained using stellar
population models with variable element abundance ratios.  Almost all
the sample bulges are characterized by a young stellar population,
on-going star formation, and a solar $\alpha/$/Fe enhancement. Their
metallicity spans from high to sub-solar values. No significant
gradient in age and $\alpha/$Fe enhancement is measured, whereas only
in a few cases a negative metallicity gradient is found.
These properties suggest that a pure dissipative collapse is not able
to explain formation of all the sample bulges and that other
phenomena, like mergers or acquisition events, need to be
invoked. Such a picture is also supported by the lack of a correlation
between the central value and gradient of the metallicity in bulges
with very low metallicity.
The stellar populations of the bulges hosted by low surface-brightness
discs share many properties with those of high surface-brightness
galaxies. Therefore, they are likely to have common formation
scenarios and evolution histories. A strong interplay between bulges
and discs is ruled out by the fact that in spite of being hosted by
discs with extremely different properties, the bulges of low and high
surface-brightness discs are remarkably similar. 

\end{abstract}


\section{Introduction}
\label{sec:introduction}

Galaxies with a central face-on surface brightness fainter than 22.6
\mas\/ in the $B$ band are classified as low surface-brightness (LSB)
systems. Although they are more difficult to be identified than high
surface-brightness (HSB) galaxies, LSB galaxies do not occupy a niche
in galactic astrophysics. Indeed, they constitute up to 50 per cent of
the galaxy population and, consequently, represent one of the major
baryonic repositories in the Universe \citep[see][for a
  review]{bothetal97}. Most of LSB are dwarf galaxies with usually
blue colour \citep{zacketal05, voroetal09}. Nevertheless, LSB galaxies
are characterized by different morphologies (ranging from dwarf
irregulars to giant spirals) and cover a wide range of colors ($0.3 <
B-V < 1.7$) suggesting that they can follow a variety of evolutionary
paths. The typical gas surface density of LSB discs is below the
critical threshold necessary for star formation, despite the fact that
they have an higher content of neutral hydrogen with respect to their
HSB counterparts \citep{vdhuetal93}. This inability to condense atomic
gas into molecular gas results in a very low star formation rate and
in a significantly slower evolution of the galaxy.

Although most of the LSB galaxies are bulgeless, there are also
galaxies with a LSB disc and a significant bulge component
\citep[][hereafter Paper I]{beijetal99, pizzetal08}. It is not known
whether these bulges are similar to those of HSB galaxies and
whether their properties do depend on the LSB nature of their
host discs.

An invaluable piece of information to understand the processes of
formation and evolution of bulges in LSB galaxies is imprinted in
their stellar populations. To this aim, the central values and radial
profiles of age, metallicity, and $\alpha$/Fe enhancement of the
stellar component can be used to test the predictions of theoretical
models, as already done for the bulges of HSB galaxies
\citep[see][]{mehletal03, rampetal05, sancetal06p,
  annietal07,jabletal07,zhao12}.
Gas dissipation toward the galaxy centre with subsequent star
formation and blowing of galactic winds produces a gradient in the
radial profile of metallicity. Therefore, a metallicity gradient is
expected in bulges formed via monolithic collapse
\citep[e.g.,][]{kawaetal01, kobayashi04}. A relationship between the
gradient steepening and galaxy mass is also expected
\citep{pipietal10} as well as a strong gradient in $\alpha$/Fe
enhancement \citep{fesi02}. However, according to \citet{pipietal08}
other processes have also to be taken into account to explain the
observed abundance ratios. In particular, the interplay between the
timescale of star formation and gas flows produces the metallicity
gradient and in the meanwhile it flattens the gradient of $\alpha$/Fe enhancement
\citep{pipietal08}.
The metallicity gradient is expected to be very shallow (or even
absent) in bulges built by merging \citep{besh99}. In fact, dry
mergers mix up the galaxy stars erasing the pre-existing population
gradients \citep{pipietal10}. The metallicity gradient is rarely
enhanced by secondary events of star formation which eventually occur
in wet mergers \citep{kobayashi04}. If this happens, a clear signature
has to be observed in the age radial profile for several Gyrs
\citep{hopketal09}.
The predictions for bulges assembled through long timescale
processes, such as dissipationless secular evolution of the disc
component, are more controversial. According to this scenario, the
bulge formed by the redistribution of disc stars
\citep{korken04, kormetal09}. The population gradients which are
eventually present in the progenitor disc could be either amplified,
since the resulting bulge has a smaller scale length than disc, or
erased as a consequence of disc heating \citep{mooretal06}.

Whereas the measurement of the emission lines in the optical spectra
of LSB galaxies has been extensively performed to derive the
metallicity of the ionized gas
\citep{impetal01,bergetal03,ramyetal11,lianetal11}, to date the
properties of the underlying stellar populations have been obtained
only in a few cases \citep{bergetal03}. To address this issue, here we
present a detailed photometric and spectroscopic study of a sample of
8 bulges hosted in LSB discs. The analysis of the spectral absorption
lines allowed us to derive the age and metallicity of the stellar
populations and estimate the efficiency and timescale of the last
episode of star formation in order to disentangle between early rapid
assembly and late slow growing of the bulge.

The paper is organized as follows. The galaxy sample is presented in
Section \ref{sec:sample}. The photometric and spectroscopic
observations together with the data reduction and analysis are
described in Section \ref{sec:data}. The properties of the stellar
populations of the bulges of the sample galaxies are investigated in
Section \ref{sec:stellarpop}. Finally, the conclusions are given in
Section \ref{sec:conclusions}.

\section{Sample selection}
\label{sec:sample}

\begin{figure*}
\centering
\includegraphics[angle=0.0,width=0.91\textwidth]{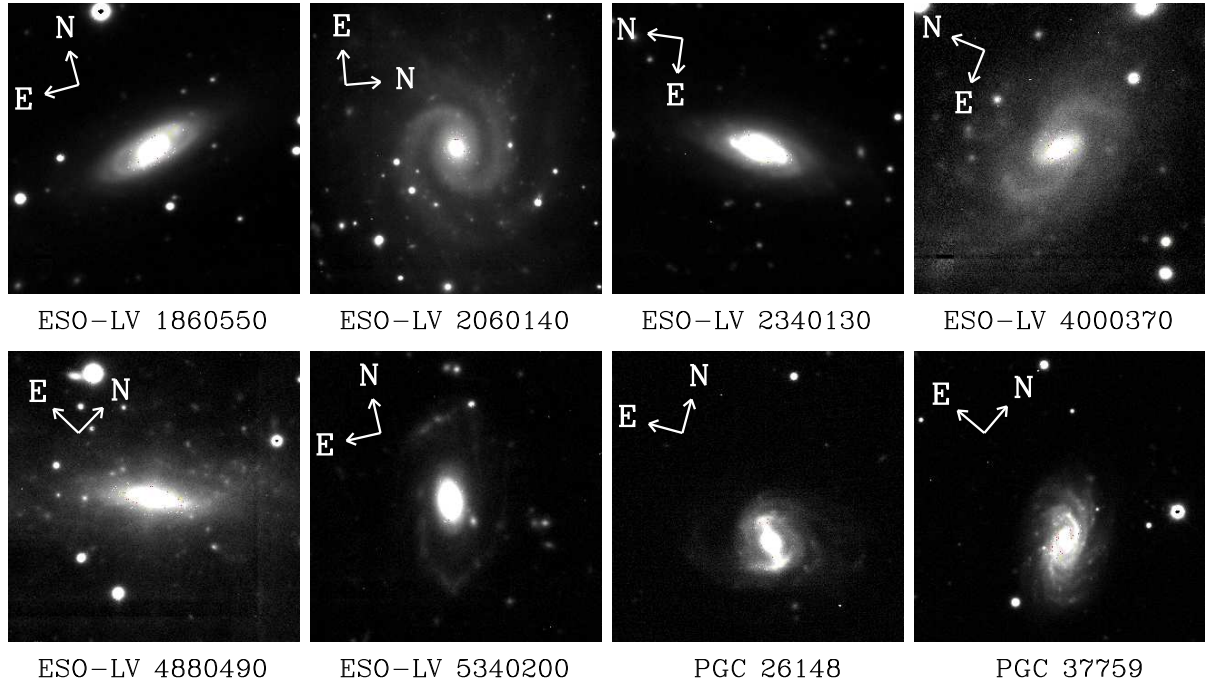}\\
\caption{Optical images of the sample galaxies. The orientation is
  specified by the arrows indicating north and east in the upper-left
  corner of each image. The size of the plotted region is about 80 arcsec $\times$
  80 arcsec.
\label{fig:images}}
\end{figure*}

All the sample galaxies were selected to be spiral galaxies with a bulge and a LSB
disc. 
Six of them were taken from Paper I and their broad-band images and
long-slit spectra are already available to us. We therefore refer to
Paper I for a detailed description of the photometric and kinematic
data we are using in this paper to measure the line-strength indices 
of these galaxies.
In addition, two more late-type spirals were taken from the catalogue
of candidate LSB galaxies by \citet{impetal96}. It should be
  noted that these galaxies have central surface brightnesses that
  place them on the low side of normal spirals, but are not examples
  of typical LSB galaxies. Nevertheless, the LSB nature of their
discs was confirmed by a detailed decomposition of the
surface-brightness distribution (see Section \ref{sec:photdec} for
details).
Therefore, the final sample is made by 8 LSB spiral galaxies with a
morphological type ranging from Sa to Sm and including some barred
galaxies. An overview of their basic properties which includes
morphological type, size, apparent and absolute magnitude, and distance
is given in Table
\ref{tab:sample}  and their optical images are shown in \ref{fig:images}.

%%%%%%%%%%%%%%%%%%%%%%%%%%%%%%%%%%%%%%%%%%%%%%%%%%%%%%%%%%%%%%%%%%%%%%%%%%%%%%%%
%% Figure 0
%%%%%%%%%%%%%%%%%%%%%%%%%%%%%%%%%%%%%%%%%%%%%%%%%%%%%%%%%%%%%%%%%%%%%%%%%%%%%%%%

%%%%%%%%%%%%%%%%%%%%%%%%%%%%%%%%%%%%%%%%%%%%%%%%%%%%%%%%%%%%%%%%%%%%%%%%%%%%%%%%
\begin{table*}
\caption{Parameters of the sample galaxies. The columns show the
  following. Column (2): morphological classification from Lyon
  Extragalactic Database (LEDA); column (3): numerical morphological
  type from LEDA. The typical error on T is 1.0; column (4): apparent isophotal diameters measured at a
  surface-brightness level of $\mu_B = 25$ mag arcsec$^{-2}$ from LEDA;
  column (5): total observed blue magnitude from LEDA; column (6):
  radial velocity with respect to the CMB reference frame from LEDA;
  column (7): distance obtained as $V_{\rm CMB}/H_0$ with $H_0= 75$ km
  s$^{-1}$ Mpc$^{-1}$; column (8): absolute total blue magnitude from
  $B_T$ corrected for extinction as in LEDA and adopting $D$; column
  (9): source of the photometric and spectroscopic data: 1 = Paper I, 2 =
  this paper.}
\begin{center}
\begin{small}
\begin{tabular}{llc cr ccc c}
\hline
\noalign{\smallskip}
\multicolumn{1}{c}{Galaxy} &
\multicolumn{1}{c}{Type} &
\multicolumn{1}{c}{$T$} &
\multicolumn{1}{c}{$D_{25}\,\times\,d_{25}$} &
\multicolumn{1}{c}{$B_{\rm T}$} &
\multicolumn{1}{c}{$V_{\rm CMB}$} &
\multicolumn{1}{c}{$D$} &
\multicolumn{1}{c}{$M_{B_{\rm T}}$} &
\multicolumn{1}{c}{Source}\\ 
\noalign{\smallskip}
\multicolumn{1}{c}{} &
\multicolumn{1}{c}{} &
\multicolumn{1}{c}{} &
\multicolumn{1}{c}{(arcmin)} &
\multicolumn{1}{c}{(mag)} &
\multicolumn{1}{c}{(\kms)} &
\multicolumn{1}{c}{(Mpc)} &
\multicolumn{1}{c}{(mag)} &
\multicolumn{1}{c}{}\\
\noalign{\smallskip}
\multicolumn{1}{c}{(1)} &
\multicolumn{1}{c}{(2)} &
\multicolumn{1}{c}{(3)} &
\multicolumn{1}{c}{(4)} &
\multicolumn{1}{c}{(5)} &
\multicolumn{1}{c}{(6)} &
\multicolumn{1}{c}{(7)} &
\multicolumn{1}{c}{(8)} &
\multicolumn{1}{c}{(9)}\\
\noalign{\smallskip}
\hline
\noalign{\smallskip}  
ESO-LV~1860550 & Sab        &$2.0$& $1.7\times0.6$ &13.96 &  4594 & 60.1 & $-19.93$ &1\\
ESO-LV~2060140 & SABc       &$5.0$& $1.6\times0.8$ &14.89 &  4672 & 60.5 & $-19.01$ &1\\
ESO-LV~2340130 & Sbc        &$3.9$& $1.4\times0.5$ &14.68 &  4644 & 60.9 & $-19.24$ &1\\
ESO-LV~4000370 & SBc        &$5.9$& $1.8\times0.9$ &14.47 &  2876 & 37.5 & $-18.40$ &1\\
ESO-LV~4880490 & SBd        &$7.9$& $1.9\times0.6$ &15.03 &  1870 & 25.0 & $-16.95$ &1\\
ESO-LV~5340200 & Sb         &$3.1$& $0.9\times0.3$ &16.19 & 17446 & 226.7& $-20.58$ &1\\
PGC~26148      & Sm         &$9.0$& $0.7\times0.5$ &15.71 & 11718 & 156.2& $-20.25$ &2\\
PGC~37759      & Sc         &$6.0$& $0.6\times0.4$ &15.89 & 14495 & 193.2& $-20.54$ &2\\
\noalign{\smallskip}
\hline
\noalign{\medskip}
\end{tabular}
\end{small}
\label{tab:sample}
\end{center}
\end{table*}
%%%%%%%%%%%%%%%%%%%%%%%%%%%%%%%%%%%%%%%%%%%%%%%%%%%%%%%%%%%%%%%%%%%%%%%%%%%%%%%%

\section{Observations, data reduction, and analysis}
\label{sec:data}

\subsection{Broad-band imaging and long-slit spectroscopy}
\label{sec:observations}

The photometric and spectroscopic observations of PGC~26148 and
PGC~37759 were carried out with the Very Large
Telescope (VLT) of the European Southern Observatory (ESO) at Paranal
Observatory on 2006 March 23-25 and 2008 February 17-18.
The two galaxies were observed with the UT1 unit mounting the
FOcal Reducer and low dispersion Spectrograph 2 (FORS2). The detector
consists of a mosaic of 2 MIT/LL CCDID-20 CCDs separated by a gap of
480 $\mu$m along the spatial direction. Each CCD has
$2048\,\times\,4096$ pixels of $15\,\times\,15$ $\mu$m$^2$. A
$2\,\times\,2$ pixel binning was adopted giving a spatial scale of
$0.25$ arcsec pixel$^{-1}$ with a field of view of $6.8\,\times\,6.8$
arcmin$^2$. The gain and readout noise were set to 1.43 $e^-$
ADU$^{-1}$ and 2.90 $e^-$ (rms), respectively.

For each galaxy three 20-s broad-band images were obtained with the
R\_special$+$76 filter centred at 6550 \AA\ with a FWHM of 1650 \AA .

The long-slit spectra were taken using the grism GRIS\_1400V+18 with
1400 $\rm grooves\,mm^{-1}$ in combination with the 0.7 arcsec
$\times$ 6.8 arcmin slit. The wavelength range between 4560 and 5840
\AA\ was covered with a reciprocal dispersion of 0.64
\AA\ pixel$^{-1}$ after pixel binning. This set up guarantees an
adequate oversampling of the instrumental broadening function. Indeed,
the instrumental velocity resolution was $1.31$ \AA\ (FWHM)
corresponding to $\sigma_{\rm inst}=33$ \kms\ at 5000 \AA . It was
estimated by measuring the width of the emission lines of a comparison
arc spectrum after the wavelength calibration.

Major and minor-axis spectra were obtained for PGC~37759 ($\rm
P.A._{\rm MJ} = 40^\circ$, $\rm P.A._{\rm MN} = 130^\circ$), whereas only
major-axis spectra were taken for PGC~26148 ($\rm P.A._{\rm MJ} =
15^\circ$). For each
position three 40-minutes spectra were taken. The position of the slit
centre along the spatial direction was slightly drifted so that the
gap between the two CCDs was imaged in different locations.

The standard calibration frames for imaging and spectroscopy (i.e.,
darks, biases, and flatfields) as well as the spectra of the
comparison arc lamp were taken in the afternoon or at the twilight
before each observing night. Spectrophotometric standards were
observed to ensure the flux calibration of the spectra.
The typical value of the seeing FWHM during the observing runs was
$1.2$ arcsec as measured by fitting a two-dimensional Gaussian to the
field stars in the acquisition images.

\subsection{Data reduction}
\label{sec:reduction}

The data reduction of the images and spectra was performed standard
IRAF\footnote{IRAF is distributed by the National Optical Astronomy
  Observatories which are operated by the Association of Universities
  for Research in Astronomy under cooperative agreement with the
  National Science Foundation.} routines.

After dark and bias subtraction, all the images were corrected for
pixel-to-pixel intensity variations by using an average sky flatfield
frame for each night.
The different images of each galaxy were shifted and aligned
to an accuracy of a few hundredths of pixel using common field stars
as reference. After checking that their point spread functions (PSFs)
were comparable within a few percent, the images were combined to
obtain a single image.
The cosmic rays were identified during the combination process and
removed using a sigma-clipping rejection algorithm.
The sky background level was removed from the combined images by
fitting a second-order polynomial to the regions free of sources.  The
median value of the residual sky level was determined in a large
number of $5\times5$ pixel areas. These areas were selected in empty
regions of the images, free of objects and far from the galaxy. The
mean of these median values was zero, as expected. For the error in
the sky determination we adopted half of the difference between the
maximum and minimum of the median values obtained in the sampled
areas.
For each combined image the photometric zero-point was derived using
the Sloan Digital Sky Survey (SDSS) $r$-band magnitude of several
stars present in the field of view. The total $r$-band magnitude of
each galaxy is consistent within 0.1 mag with the value quoted in SDSS
Data Release 7 \citep{DR7}.

After bias subtraction, the flatfield correction of the spectra
was performed by means of quartz lamp spectra. They were normalised
and divided into all the galaxy and star spectra, to correct for
pixel-to-pixel sensitivity variations and large-scale illumination
patterns due to slit vignetting.
The cosmic rays were identified by comparing the photon
counts in each pixel with the local mean and standard deviation and
eliminated by interpolating over a suitable value. The residual cosmic
rays were eliminated by manually editing the spectra.
Each spectrum was rebinned using the wavelength solution obtained from
the corresponding arc-lamp spectrum. We checked that the wavelength
rebinning had been done properly by measuring the difference between
the measured and predicted wavelengths of about 20 of unblended
arc-lamp lines which were distributed over the whole spectral range of
a wavelength-calibrated spectrum. The resulting accuracy in the
wavelength calibration is better than 4 \kms.
All the spectra were corrected for CCD misalignment and the flux
calibrated using the sensitivity function from the spectrophotometric
star spectrum of the corresponding night.
The spectra obtained for the same galaxy in the same position were
co-added using the centre of the stellar continuum as reference. This
allowed to improve the signal-to-noise ratio ($S/N$) of the resulting
two-dimensional spectrum. In such a spectrum, the contribution of the
sky was determined by interpolating a one-degree polynomium along the
outermost 20 arcsec at the two edges of the slit, where the galaxy
light was negligible, and then subtracted. A sky subtraction better
than 1 per cent was achieved.

\subsection{Two-dimensional photometric decomposition}    
\label{sec:photdec}

The photometric decomposition of the $r$-band images of PGC~26148, and
PGC~37759 was performed using the Galaxy Surface Photometry
Two-Dimensional Decomposition (GASP2D) algorithm by
\citet{mendetal08}. The structural parameters of the galaxies were
derived assuming the galaxy surface-brightness distribution to be the
sum of the contributions of a S\'ersic bulge, an exponential disc,
and, if necessary, a Ferrers bar.  Conventional bulge-disc
decompositions using GASP2D have been explored in detail in
\cite{pizzetal08} and \cite{moreetal08}. For sake of clarity, here we
briefly describe the main properties of the bar component adopted in
GASP2D.

We adopted the projected surface density of  a three-dimensional Ferrers
ellipsoid  \citep{Ferrers1877}   to  describe  the  surface-brightness
profile of the bars

%-------------------------------------------------------------------
\begin{equation}
I_{\rm bar}(r)=I_{\rm 0,bar}\left[1-\left(\frac{r_{\rm bar}}{a_{\rm bar}}\right)2\right]^{n_{\rm bar}+0.5}
\qquad r_{\rm bar} \le a_{\rm bar},
\end{equation}
%-------------------------------------------------------------------
%
where $I_{\rm 0,bar}$, $a_{\rm bar}$, and $n_{\rm bar}$ represent the
central surface-brightness, length, and shape parameter of bar,
respectively. Due to the high degree of degeneracy that $n_{\rm bar}$
introduces in the fit, we decided to keep it as a fixed parameter
($n_{\rm bar}=2$ \citet{lauretal05}) during the fitting process.  All
the bar models were built up in a frame of generalized ellipses
\citep{Athanassoula90}.  Thus, the bar reference system is defined as

%---------------------------------------------------------------------
\begin{eqnarray}
r_{\rm bar}&=&\left[(-(x-x_0) \sin{{\rm PA_{bar}}} + (y-y_0)
\cos{{\rm PA_{bar}}})^c  \right. \nonumber \\ & & - \left.((x-x_0)
\cos{{\rm PA_{bar}}} + (y-y_0) \sin{{\rm PA_{bar}}})^c  \right. \nonumber \\ & & / \left. 
q_{\rm bar}^{\,c}\right]^{1/c},
\label{eqn:bar_radius}
\end{eqnarray}
%---------------------------------------------------------------------
%
where $q_{\rm bar}$ and PA$_{\rm bar}$ are the axis ratio and position
angle of the  bar, respectively. The parameter $c$  controls the shape
of the  isophotes. A value of  $c=2$ corresponds to  a perfect ellipse,
$c>2$ to a boxy shape and  $c<2$ to a discy shape \citep[see][for more
  details]{agueetal09}.

To derive the photometric parameters of the bulge (effective surface
brightness $I_{\rm e}$, effective radius $r_{\rm e}$, shape parameter
$n$, major-axis position angle PA$_{\rm b}$, and axial ratio $q_{\rm
  b}$) and disc (central surface brightness $I_0$, scalelength $h$,
major-axis position angle PA$_{\rm d}$, and axial ratio $q_{\rm d}$)
and the position of the galaxy centre $(x_0,y_0)$ we fitted
iteratively a model of the surface brightness $I_{\rm m}(x,y) = I_{\rm
  b}(x,y) + I_{\rm d}(x,y)$ to the pixels of the galaxy image, using a
non-linear least-squares minimisation based on the robust
Levenberg-Marquardt method by \citet{moreetal80}.  The actual
computation was done using the MPFIT \citep{Markwardt09} under the IDL\footnote{The
  Interactive Data Language is distributed by ITT Visual Information
  Solutions.} environment.
Each image pixel was weighted according to the variance of its
total observed photon counts due to the contribution of both the galaxy
and sky, and determined assuming photon noise limitation and taking
into account for the detector read-out noise.
The seeing effects were taken into account by convolving the model
image with a circular Gaussian PSF with the FWHM measured from the
field stars in the galaxy image. The convolution was performed as a
product in Fourier domain before the least-squares minimisation.
  
%%%%%%%%%%%%%%%%%%%%%%%%%%%%%%%%%%%%%%%%%%%%%%%%%%%%%%%%%%%%%%%%%%%%%%%%%%%%%%%%
%% Figure 1
\begin{figure*}
\centering
\includegraphics[angle=0.0,width=0.81\textwidth]{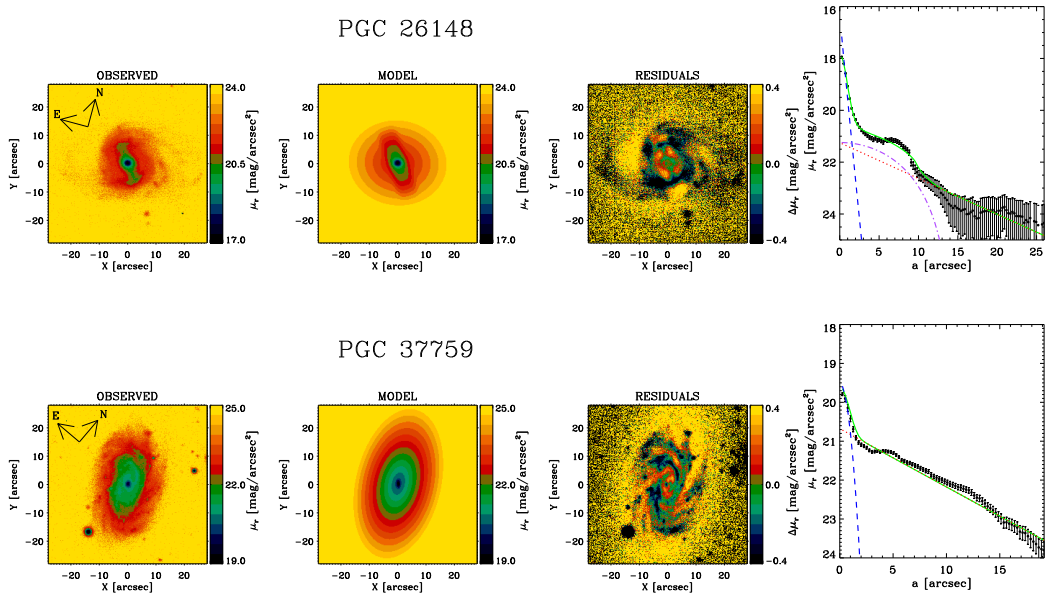}\\
\caption{ Two-dimensional photometric decomposition of PGC~26148 (top panel)
  and PGC~37759 (bottom panel). The FORS2 $r$-band image, best-fit image,
  residual (i.e., observed-model) image, and surface-brightness radial
  profile are shown (from left to right panel). In the right-hand panel the
  ellipse-averaged radial profile of the surface brightness measured
  in the FORS2 (dots) and model image (green continuous line) are
  shown. The dashed blue, dash-dotted purple line, and dotted
  red lines represent the intrinsic surface-brightness radial profiles
  of the bulge, bar, and disc, respectively.
\label{fig:decomposition}}
\end{figure*}
%%%%%%%%%%%%%%%%%%%%%%%%%%%%%%%%%%%%%%%%%%%%%%%%%%%%%%%%%%%%%%%%%%%%%%%%%%%%%%%%
\begin{table*}
\caption{Photometric parameters of the bulge and disc of PGC~26148 and
  PGC~37759. The columns show the following. 
  Column (2) the effective surface brightness of the bulge; 
  column (3): effective radius of the bulge; 
  column (4): shape parameter of the bulge;
  column (5): axial ratio of the bulge isophotes; 
  column (6): position angle of the bulge major axis; 
  column (7): central surface brightness of the disc; 
  column (8): scale length of the disc; 
  column (9): axial ratio of the disc isophotes; 
  column (10): position angle of the disc major axis; 
  column (11): Central surface brightness of the bar; 
  column (12): radius of the bar; 
  column (13): shape parameter of the bar; 
  column (14): axial ratio of the bar isophotes; 
  column (15): bulge-to-total luminosity ratio.
}
\begin{tiny}
\label{tab:parameters} 
\begin{tabular}{lrrrrrrrrrr}
\hline
\noalign{\smallskip}
\multicolumn{1}{c}{Galaxy} &
\multicolumn{1}{c}{$\mu_{\rm e}$} &
\multicolumn{1}{c}{$r_{\rm e}$} &
\multicolumn{1}{c}{$n$} &
\multicolumn{1}{c}{$q_{\rm b}$} &
\multicolumn{1}{c}{PA$_{\rm b}$} &
\multicolumn{1}{c}{$\mu_0$} &
\multicolumn{1}{c}{$h$} &
\multicolumn{1}{c}{$q_{\rm d}$} &
\multicolumn{1}{c}{PA$_{\rm d}$}  \\
\multicolumn{1}{c}{ } &
\multicolumn{1}{c}{(mag/arcsec$^2$)} &
\multicolumn{1}{c}{(arcsec)} &
\multicolumn{1}{c}{} &
\multicolumn{1}{c}{} &
\multicolumn{1}{c}{($^{\circ}$)} &
\multicolumn{1}{c}{(mag/arcsec$^2$)} &
\multicolumn{1}{c}{(arcsec)} &
\multicolumn{1}{c}{} &
\multicolumn{1}{c}{($^{\circ}$)} \\
\multicolumn{1}{c}{(1)} &
\multicolumn{1}{c}{(2)} &
\multicolumn{1}{c}{(3)} &
\multicolumn{1}{c}{(4)} &
\multicolumn{1}{c}{(5)} &
\multicolumn{1}{c}{(6)} &
\multicolumn{1}{c}{(7)} &
\multicolumn{1}{c}{(8)} &
\multicolumn{1}{c}{(9)} &
\multicolumn{1}{c}{(10)} \\
\noalign{\smallskip}
\multicolumn{1}{c}{} &
\multicolumn{1}{c}{} &
\multicolumn{1}{c}{} &
\multicolumn{1}{c}{} &
\multicolumn{1}{c}{} &
\multicolumn{1}{c}{$\mu_{\rm bar}$} &
\multicolumn{1}{c}{$r_{\rm bar}$} &
\multicolumn{1}{c}{$n_{\rm bar}$} &
\multicolumn{1}{c}{$q_{\rm bar}$} &
\multicolumn{1}{c}{$B/T$}  \\
\multicolumn{1}{c}{ } &
\multicolumn{1}{c}{} &
\multicolumn{1}{c}{} &
\multicolumn{1}{c}{} &
\multicolumn{1}{c}{} &
\multicolumn{1}{c}{(mag/arcsec$^2$)} &
\multicolumn{1}{c}{(arcsec)} &
\multicolumn{1}{c}{} &
\multicolumn{1}{c}{} &
\multicolumn{1}{c}{} \\
\multicolumn{1}{c}{} &
\multicolumn{1}{c}{} &
\multicolumn{1}{c}{} &
\multicolumn{1}{c}{} &
\multicolumn{1}{c}{} &
\multicolumn{1}{c}{11} &
\multicolumn{1}{c}{(12)} &
\multicolumn{1}{c}{(13)} &
\multicolumn{1}{c}{(14)} &
\multicolumn{1}{c}{(15)} \\
\noalign{\smallskip}
\hline
\noalign{\smallskip}
PGC~26148 & $18.19 \pm 0.29$ & $ 0.56 \pm 0.10$  & $ 1.04 \pm 0.05$ & $0.81 \pm 0.01$ & $ 96.4 \pm 2.0$  & $21.25 \pm 0.03$ & $  7.86 \pm 0.16$ & $0.86 \pm 0.02$ & $ 98.3 \pm 5.0$\\
& & & & &  $21.24 \pm 0.30  $ & $ 14.64 \pm 0.37 $ & $2.00$ & $ 0.35\pm 0.04 $ & $ 0.11 $\\
\noalign{\smallskip}
\noalign{\smallskip}
PGC~37759 & $20.23 \pm0.28$ & $ 0.76 \pm 0.11$  & $  0.50 \pm 0.05$ & $0.36 \pm 0.03$ & $  44.2 \pm 2.0$  & $20.68 \pm0.20$ & $  7.24 \pm 0.58$ & $0.51 \pm 0.03$ & $ 27.3 \pm 5.0$ \\
 & & & & &   $ - $ & $-$ & $- $ & $- $ & $ 0.02 $\\
\noalign{\smallskip}
\hline
\noalign{\bigskip}
\label{tab:phot_para}
\end{tabular}
\end{tiny}
\end{table*}
%%%%%%%%%%%%%%%%%%%%%%%%%%%%%%%%%%%%%%%%%%%%%%%%%%%%%%%%%%%%%%%%%%%%%%%%%%%%%%%%
\noindent
The parameters derived for the structural components of the sample
galaxies are collected in Table \ref{tab:parameters}.  The result of
the photometric decomposition of the surface-brightness distribution
of the analyzed galaxies is shown in Fig. \ref{fig:decomposition}.

Since the formal errors obtained from the $\chi^2$ minimisation are
not representative of the real errors in the structural parameters, a
series of Monte Carlo simulations was performed to derive a reliable
estimation of the errors. It was generated a set of barred and
unbarred galaxies with a total $r$-band magnitude within $14.5 \leq
M_{\rm T} \leq 15.5$ mag. The structural parameters of the artificial
galaxies were randomly chosen in the ranges
\begin{equation}
0.7 \leq r_{\rm e} \leq 17~{\rm kpc};\ 0.3 \leq q_{\rm b} \leq 0.7;\ 0.5 \leq n \leq 12.5;
\end{equation}
for the S\'ersic bulges, 
\begin{equation}
5 \leq h\leq 7~{\rm kpc};\ 0.5 \leq q_{\rm d} \leq 0.8;
\end{equation}
for the exponential discs, and 
\begin{equation}
0.5 \leq r_{\rm bar} \leq 10~{\rm kpc}; 0.2 \leq q_{\rm bar} \leq 0.7 {\rm ~with~} q_{\rm bar} \leq q_{\rm d} \leq q_{\rm b};
\end{equation}
for the Ferrers bar.
The artificial galaxies were assumed to be observed at a distance of
194 Mpc taking into account for resolution effects. Finally, a
background level and photon noise were added to the simulated images
in order to mimic the instrumental setup and $S/N$ of the actual
observations.  The artificial and observed galaxies were divided in
bins of S\'ersic index of width $\Delta n=2$.  The relative errors in
the fitted parameters of the artificial galaxies were estimated by
comparing the input and output values and were assumed to be normally
distributed. The standard deviation was adopted as the typical error
in the relevant parameter for the bulge-disc decomposition of the
observed galaxies (Table \ref{tab:parameters}).

The face-on central surface brightness of the disc in the $B$ band
$\mu^0_{0,B}$ was obtained from the $r$-band $\mu_0$ (Table
\ref{tab:parameters}) by correcting for galaxy inclination and
adopting a $B-r$ color which was computed from $g-r$ measured on the
SDSS images at $r \ge r_{\rm e}$ and transformed to $B$ band according
to \citet{chogas08}. It is $B-r=1.20$ for PGC~26148 and $B-r=1.21$ for
PGC~37759, and therefore $\mu^0_{0,B}=22.61$ and $\mu^0_{0,B}=22.62$
mag/arcsec$^2$ respectively.

\subsection{Stellar kinematics}
\label{sec:kinematics}

The stellar kinematics was measured from the galaxy absorption
features present in the wavelength range centred on the \Hb($\lambda$
4861 \AA\/) line and Mg~{\small I} line triplet ($\lambda\lambda$
5164,5173, 5184 \AA) by applying the Penalized Pixel Fitting
\citep[pPXF,][]{capems04} and Gas AND Absorption Line Fitting
\citep[GANDALF,][]{sarzetal06} IDL packages adapted for dealing with
the FORS2 spectra.
 
The galaxy spectra were rebinned along the dispersion direction to a
logarithmic scale, and along the spatial direction to obtain a
$S/N\geq40$ per resolution element. It is $S/N\simeq20$ per resolution
element at the outermost radii only. 

At each radius a linear combination of template stellar spectra from
the ELODIE library by \citet{prusou01} was convolved with the
line-of-sight velocity distribution (LOSVD) and fitted to the observed
galaxy spectrum by $\chi^2$ minimization in pixel space. The LOSVD was
assumed to be a Gaussian plus third- and fourth-order Gauss-Hermite
polynomials ${\cal H}_3$ and ${\cal H}_4$, which describe the
asymmetric and symmetric deviations of the LOSVD from a pure Gaussian
profile \citep{vdmarfran93, Gerhard93}. 
This allowed us to derive profiles of the line-of-sight velocity 
($v$), velocity dispersion ($\sigma$), and third- ($h_3$) and 
fourth-order ($h_4$) Gauss-Hermite moments of the stars. 
The spectral resolution of the template stellar spectra ($\rm
FWHM=0.5$ \AA) was degraded to match the spectral resolution of our
galaxy spectra. The stellar spectral were convolved with a Gaussian
before measuring the galaxy kinematics. Bad pixels coming from
imperfect subtraction of cosmic rays and sky emission lines were
properly masked and excluded from the fitting procedure. Ionized-gas
emission lines were simultaneously fitted as Gaussians and a low-order
additive Legendre polynomial was added to correct for the different
shape between the galaxy and template spectra. The uncertainties on
the kinematic parameters were estimated form the formal errors after
rescaling to have a reduced $\chi^2=1$.
The measured stellar kinematics is reported in Table \ref{tab:val_kin},
where the line-of-sight velocities are relative to the galaxy
centre. The folded kinematic profiles are plotted in
Fig. \ref{fig:kinplot}.
%*************************************************************

The rotation curve along the major axis of PGC~26148 is characterized by negative
values ($v\,\simeq\,-10$ km s$^{-1}$) in the innermost 1 arcsec, where
the velocity dispersion shows a decrease and $h_4$ peaks to
0.1. Further out, the rotation velocity increases to a maximum
$v\,\simeq\,20$ km s$^{-1}$ at 4 arcsec and then decreases to
zero at the last observed radius. This waving pattern in the rotation curve
has been detected in several barred galaxies \citep{betgal97} and
it is due to the presence of retrograde orbits \citep{wozpfe97}.

PGC~37759 shows a regular kinematics. The rotation velocity reaches a
maximum of about 140 km s$^{-1}$ at about 8 arcsec from the centre and
it remains constant at larger radii. No rotation is observed along the minor
axis. The velocity dispersion is about 70 km s$^{-1}$ in the centre
and decreases to 40 km s$^{-1}$ outwards. The Gauss-Hermite moments
are $h_3\,\simeq\,0$ and $h_4\,\simeq\,0$ at all radii. 

%%%%%%%%%%%%%%%%%%%%%%%%%%%%%%%%%%%%%%%%%%%%%%%%%%%%%%%%%%%%%%%%%%%%%%%%%
%%% Kinematics results

\begin{scriptsize}
\begin{figure*} 
\centering
\includegraphics[angle=90.0,width=0.431\textwidth,height=0.24\textheight]{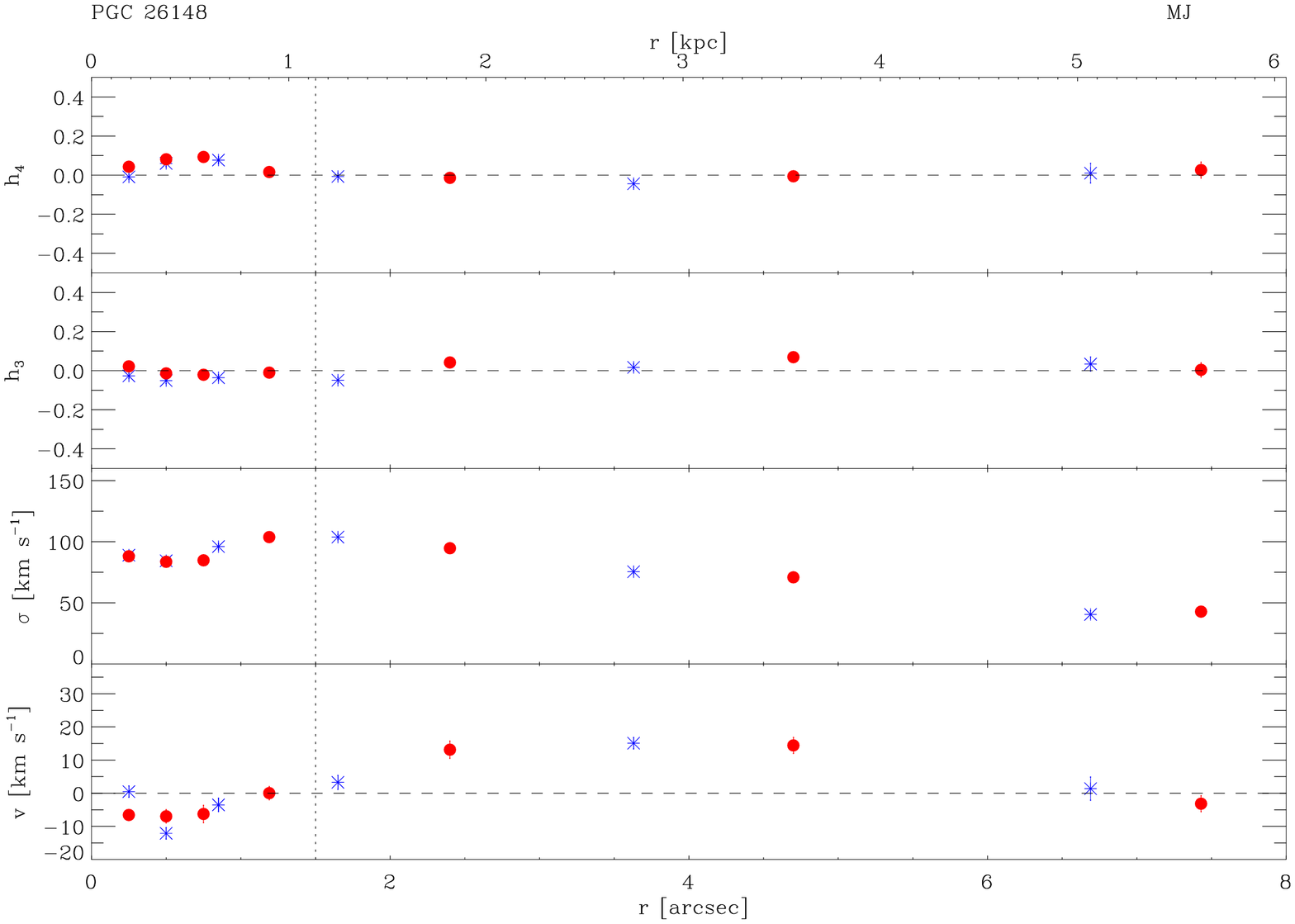}\\
\includegraphics[angle=90.0,width=0.431\textwidth,height=0.24\textheight]{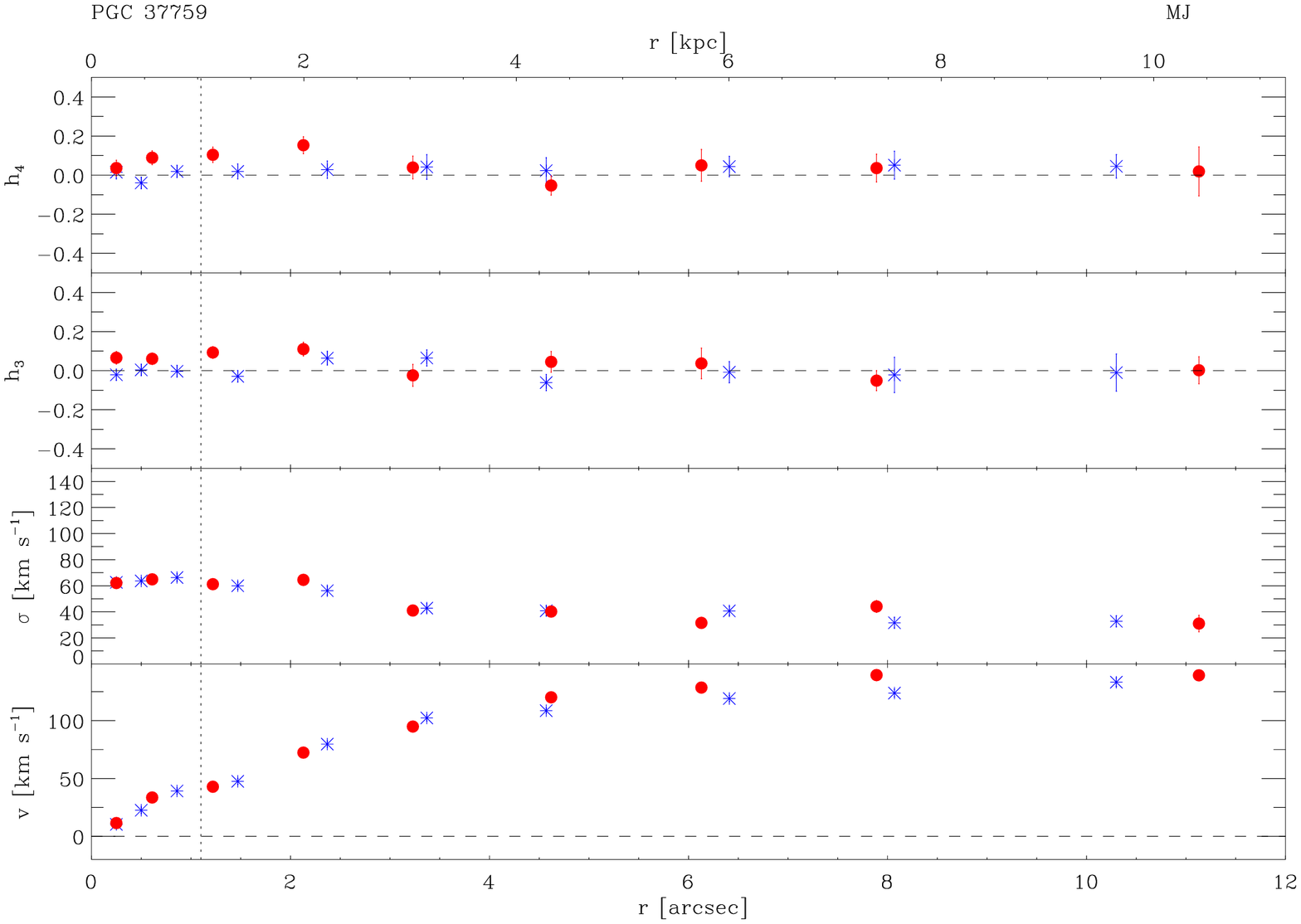}
\includegraphics[angle=90.0,width=0.431\textwidth,height=0.24\textheight]{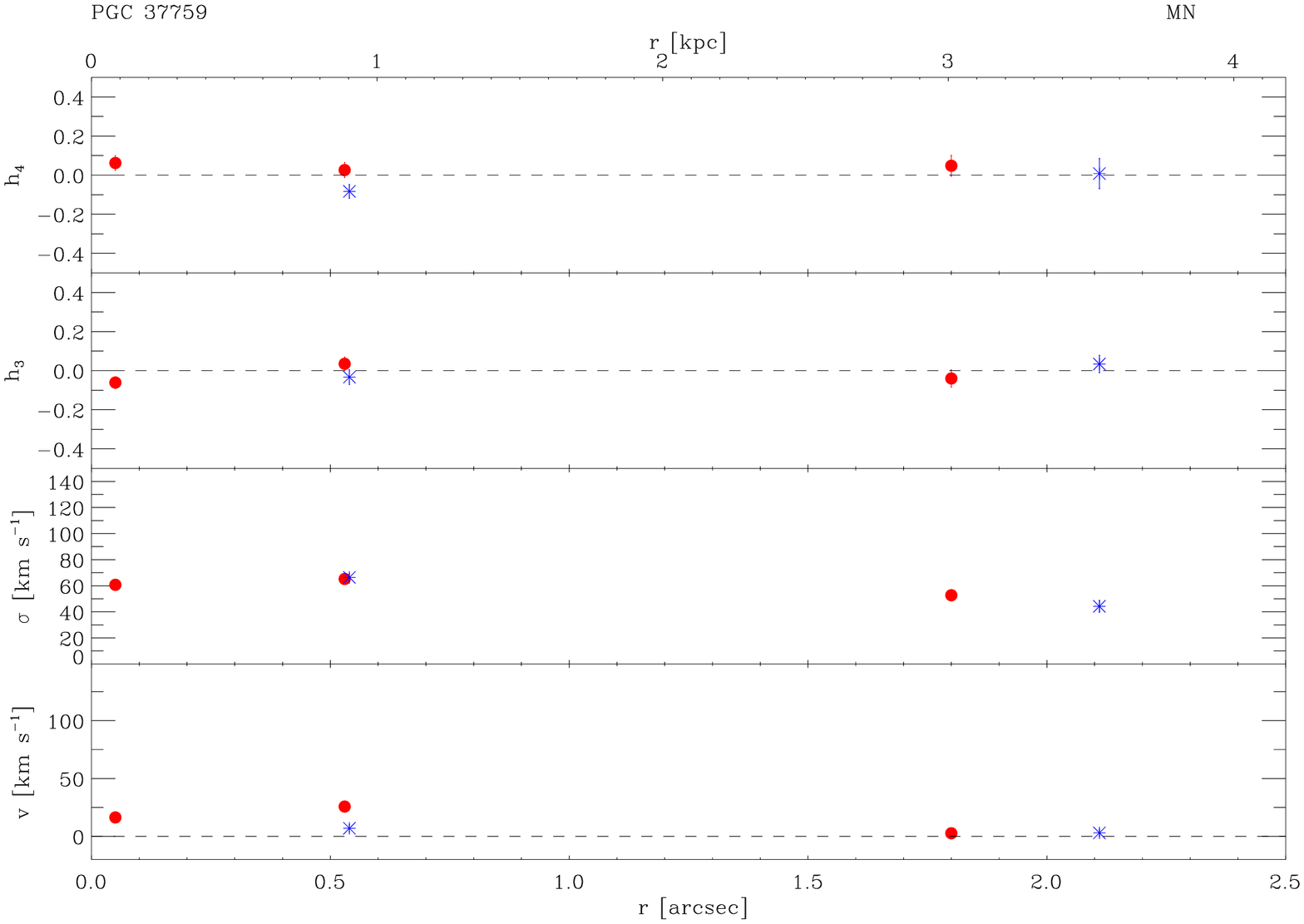}
\caption{Stellar kinematics measured along the major axis of PGC~26148
  and along the major and minor axes of PGC~37759. For each axis the
  curves are folded around the nucleus. Asterisks and dots refer to
  data measured along the approaching and residing side of the galaxy,
  respectively.  The radial profiles of the line-of-sight velocity
  ($v$) after the subtraction of the systemic velocity, velocity
  dispersion ($\sigma$), third, and fourth order coefficients of the
  Gauss-Hermite decomposition of the LOSVD ($h_3$ and $h_4$) are shown
  (from top to bottom panel). The vertical dashed line corresponds to
  the radius \Rbd\ where the surface-brightness contributions of the
  bulge and disc are equal.}
\label{fig:kinplot}
\end{figure*}
\end{scriptsize}

\subsection{Line-strength indices}
\label{sec:indices}

The Mg, Fe, and \Hb\ line-strength indices as defined by
\citet{faberetal85} and \citet{wortetal94} were measured from the flux
calibrated spectra of the 8 sample galaxies following
\citet{moreetal04, moreetal08}. The average Iron index
$\rm{\left<Fe\right> = (Fe5270 + Fe5335)/2}$ \citep{gorgetal90} and
the combined Magnesium-Iron index $[{\rm MgFe}]^{\prime}=\sqrt{{\rm
    Mg}\,b\,(0.72\times {\rm Fe5270} + 0.28\times{\rm Fe5335})}$
\citep{thmabe03} were measured too.

The difference between the spectral resolution of the galaxy spectra
and the Lick/IDS system one ($\rm FWHM = 8.4$ \AA;
\citealt{worott97}) was
taken into account by degrading our spectra through a Gaussian
convolution to match the Lick/IDS resolution before measuring the
line-strength indices. No focus correction was applied because the
atmospheric seeing was the dominant effect during observations
\citep[see][for details]{mehletal98}. The errors on the line-strength
indices were derived from photon statistics and CCD read-out noise,
and calibrated by means of Monte Carlo simulations.

The contamination of the \Hb\/ line-strength index by the \Hb\/
emission line due to the ionized gas present in the galaxy is a
problem when deriving the properties of the stellar
populations. Indeed, if the \Hb\/ emission fills the absorption line
and a proper separation of both contributions is not performed before
the analysis, the measured ages result to be artificially older. To
address this issue the \Hb\/ index was measured from the galaxy
spectrum after subtracting the contribution of the \Hb\ emission
line. Only \Hb\/ emission lines detected with a $S/N > 3$ were
subtracted from the observed spectra.

The original Lick/IDS spectra are not flux calibrated contrary to
ours. Such a difference in the continuum shape could introduce small
systematic offsets of the measured values of the indices. To derive
these offsets and to calibrate our measurements to the Lick/IDS
system, the values of the line-strength indices measured for a sample
of templates were compared to those obtained by \citet{wortetal94}.
The spectra of the template stars were obtained with the same setup as
our galaxy spectra and were retrieved from ESO Science Archive.  The
offsets were evaluated as the mean of the differences between our and
Lick/IDS line-strength values. They were neglected being smaller than
the mean error of the differences. Therefore, no offset correction was
applied to our line-strength measurements.

The measured values of \Hb , \MgFe, \Fe, \Mgb, and \Mgd\ for all the
sample galaxies are listed in Table \ref{tab:val_ind}2 and plotted in
Fig. \ref{fig:indices}.

%%%%%%%%%%%%%%%%%%%%%%%%%%%%%%%%%%%%%%%%%%%%%%%%%%%%%%%%%%%%%%%%%%%%%%%%%%%%%%%%
%% Figure 
\begin{figure*}
\centering
\includegraphics[angle=90.0,width=0.431\textwidth]{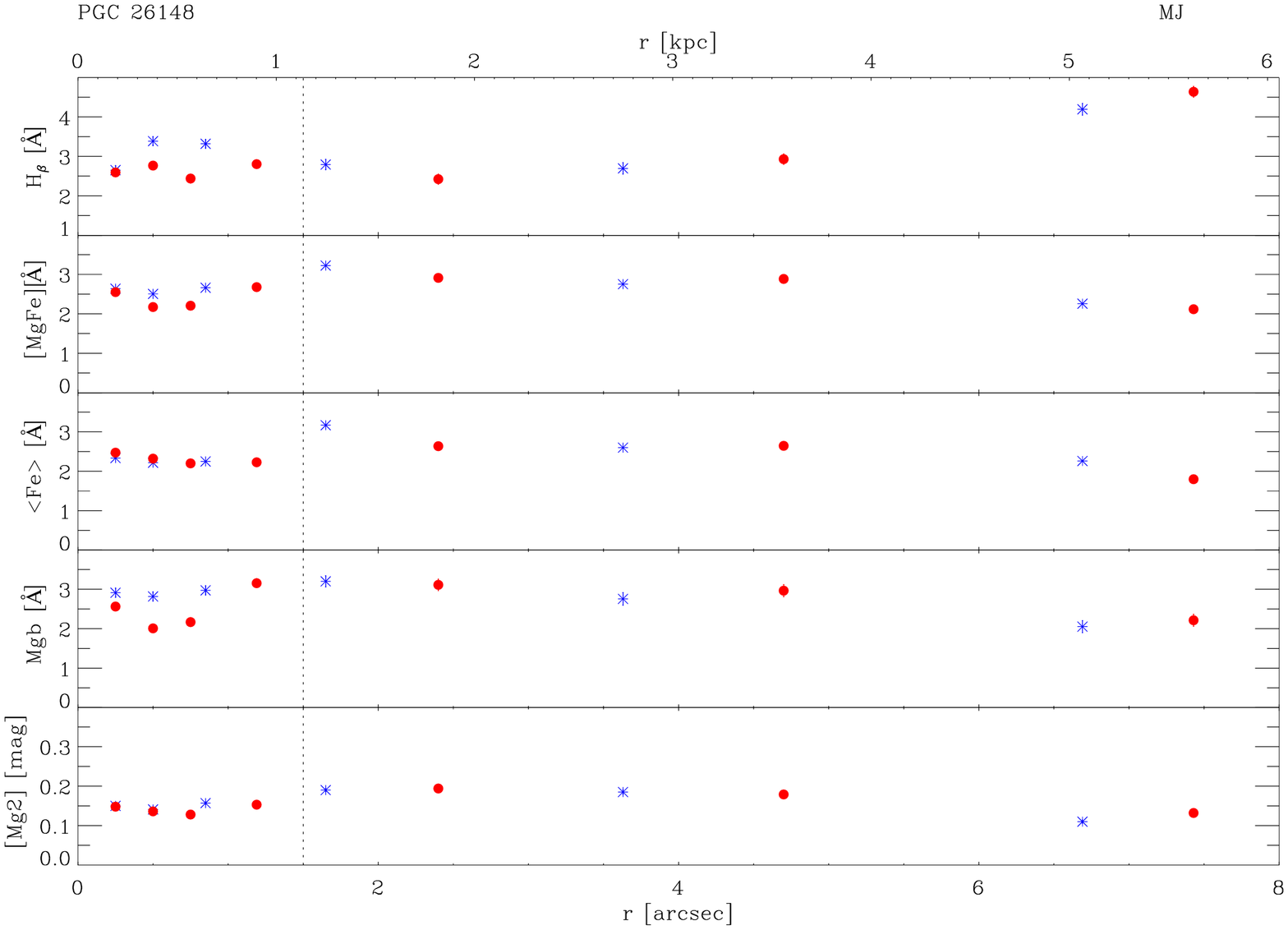}\\
\includegraphics[angle=90.0,width=0.431\textwidth]{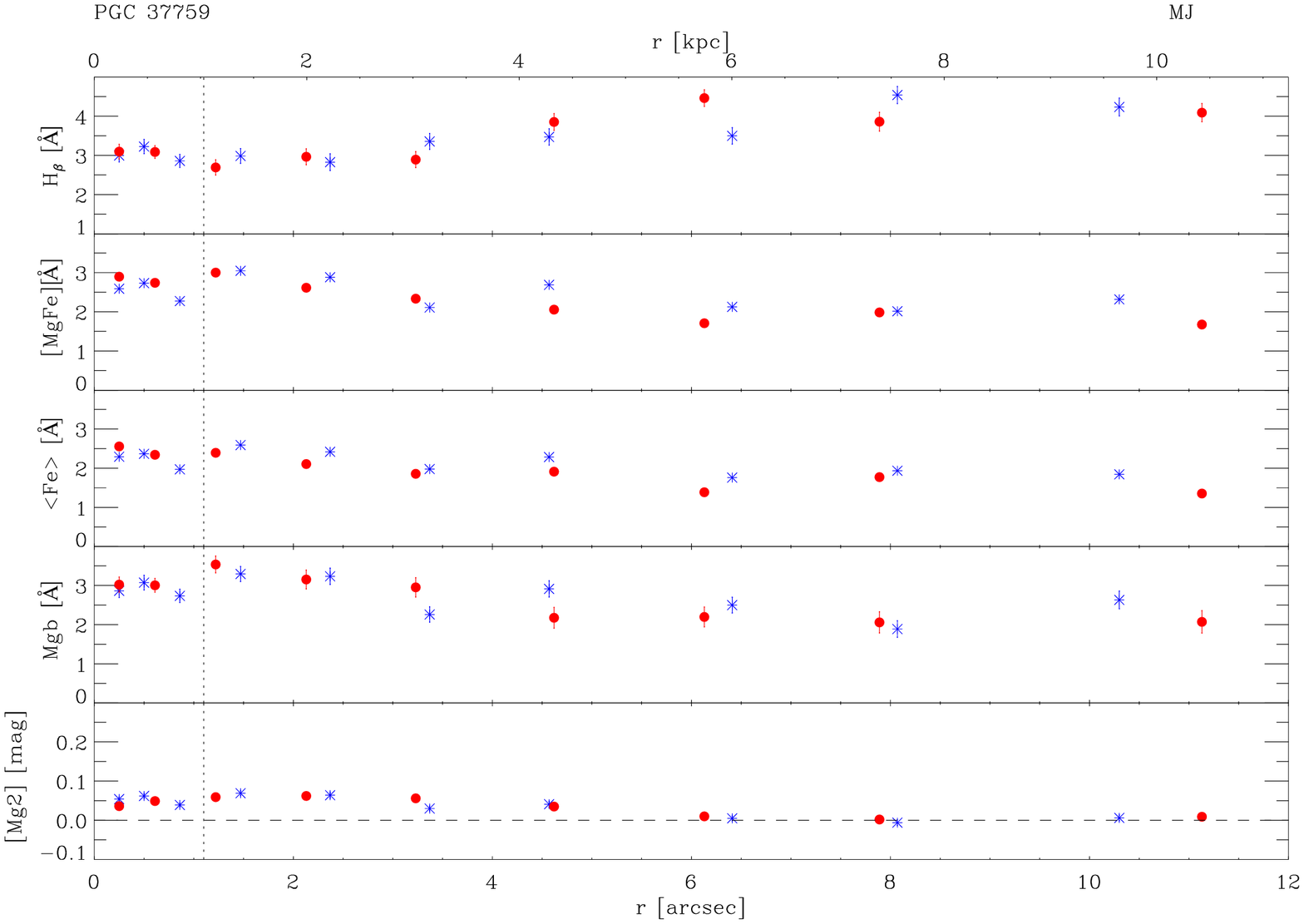}
\includegraphics[angle=90.0,width=0.431\textwidth]{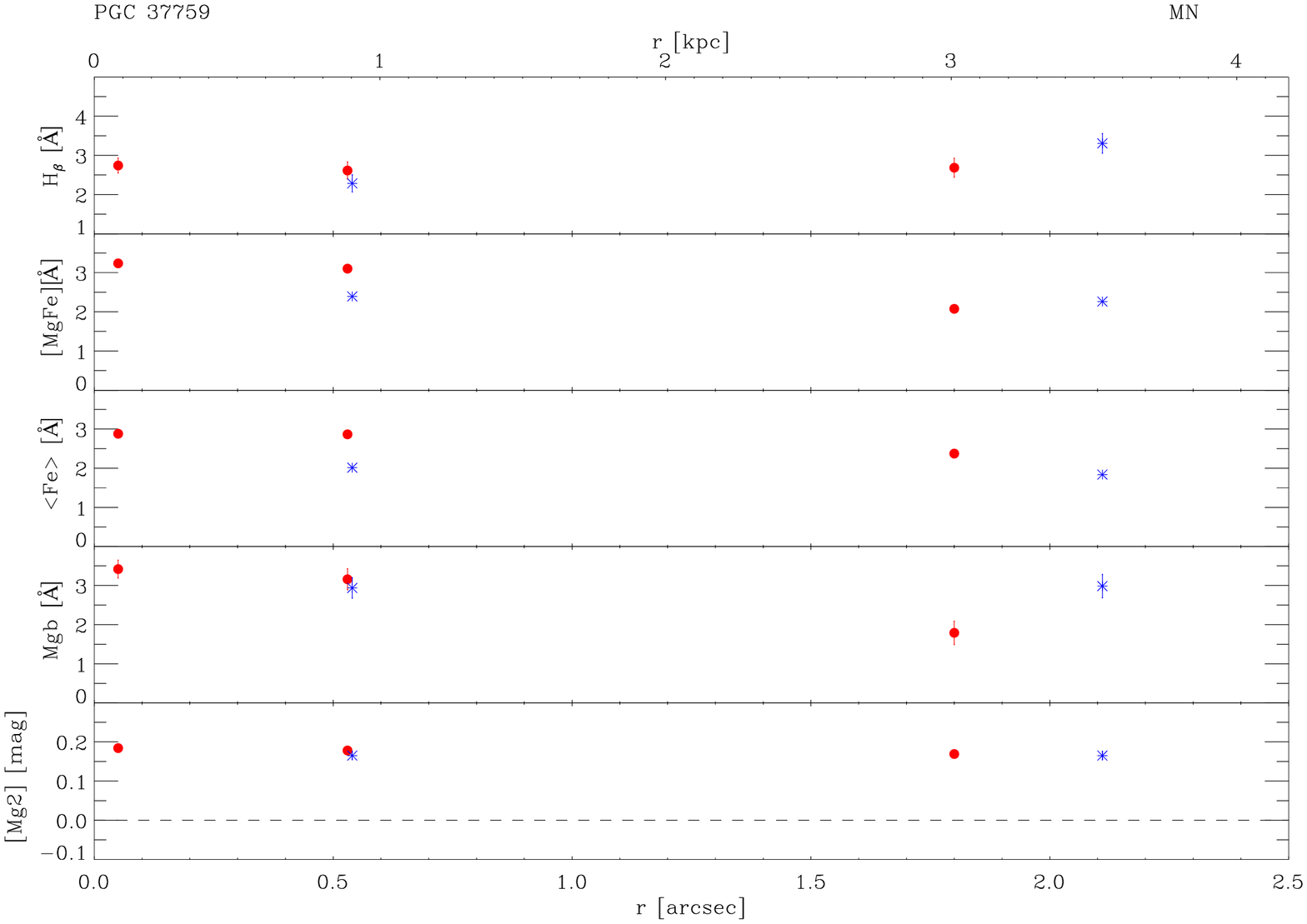}
\includegraphics[angle=90.0,width=0.431\textwidth]{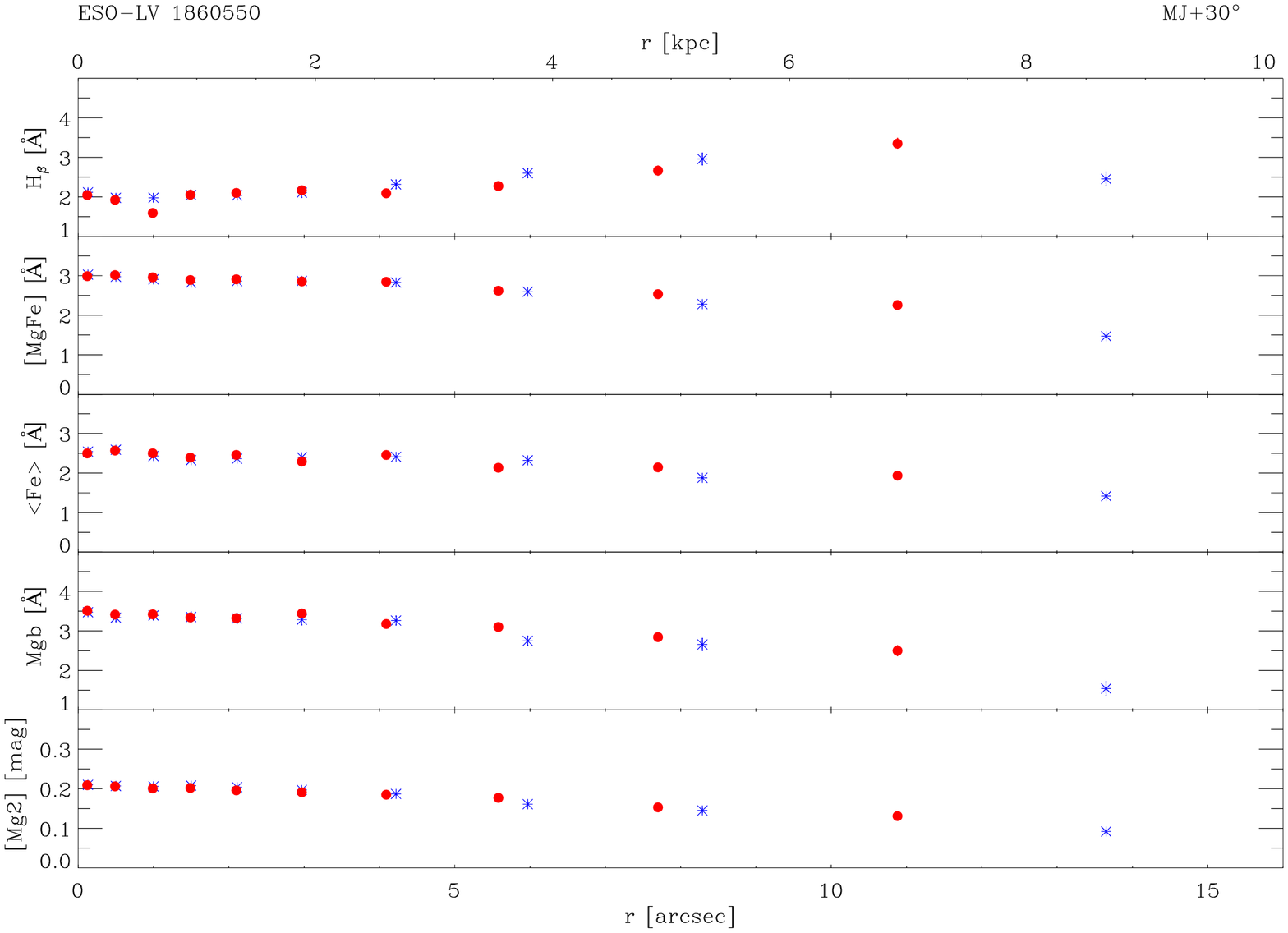}
\includegraphics[angle=90.0,width=0.431\textwidth]{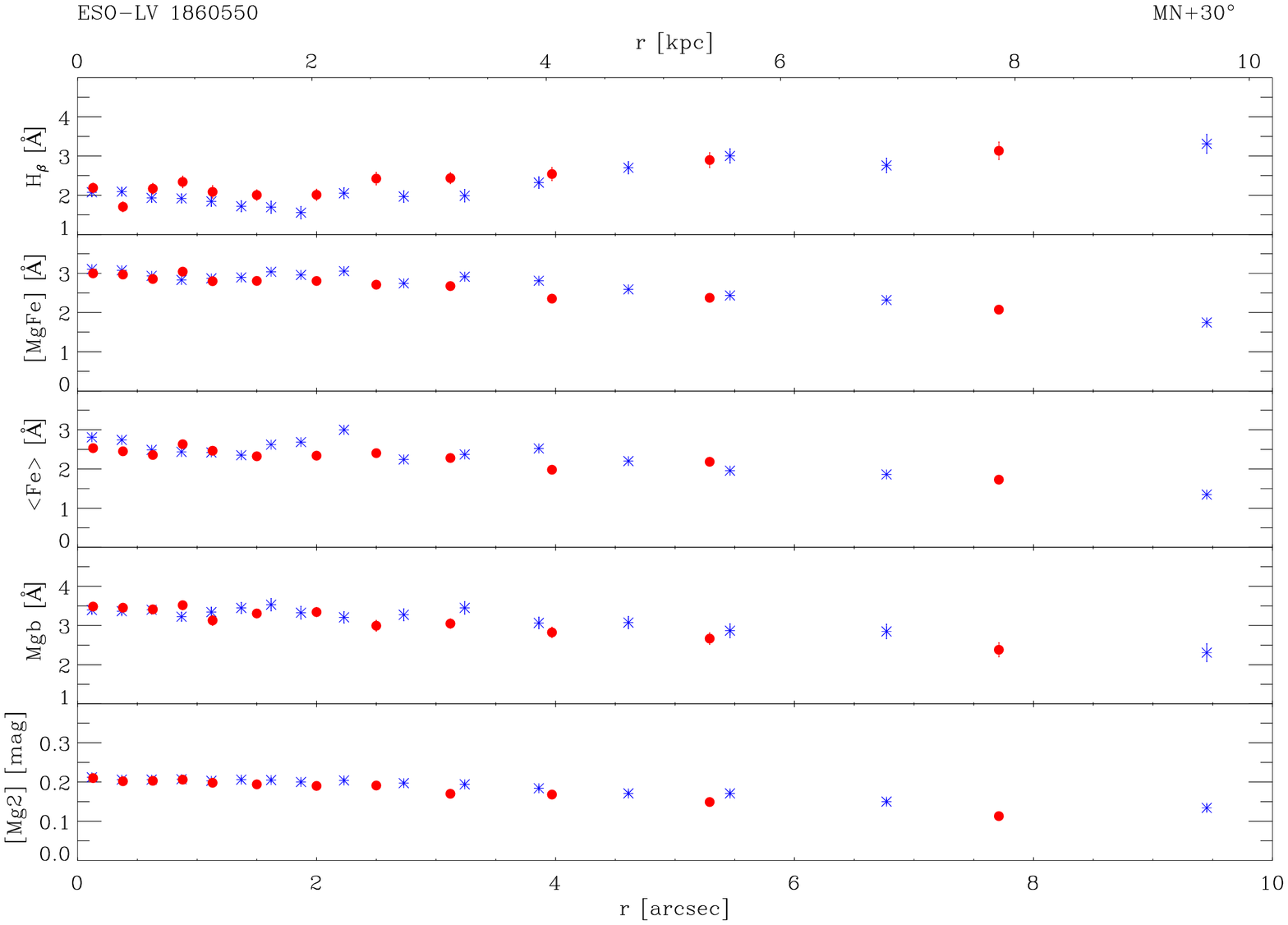}
\includegraphics[angle=90.0,width=0.431\textwidth]{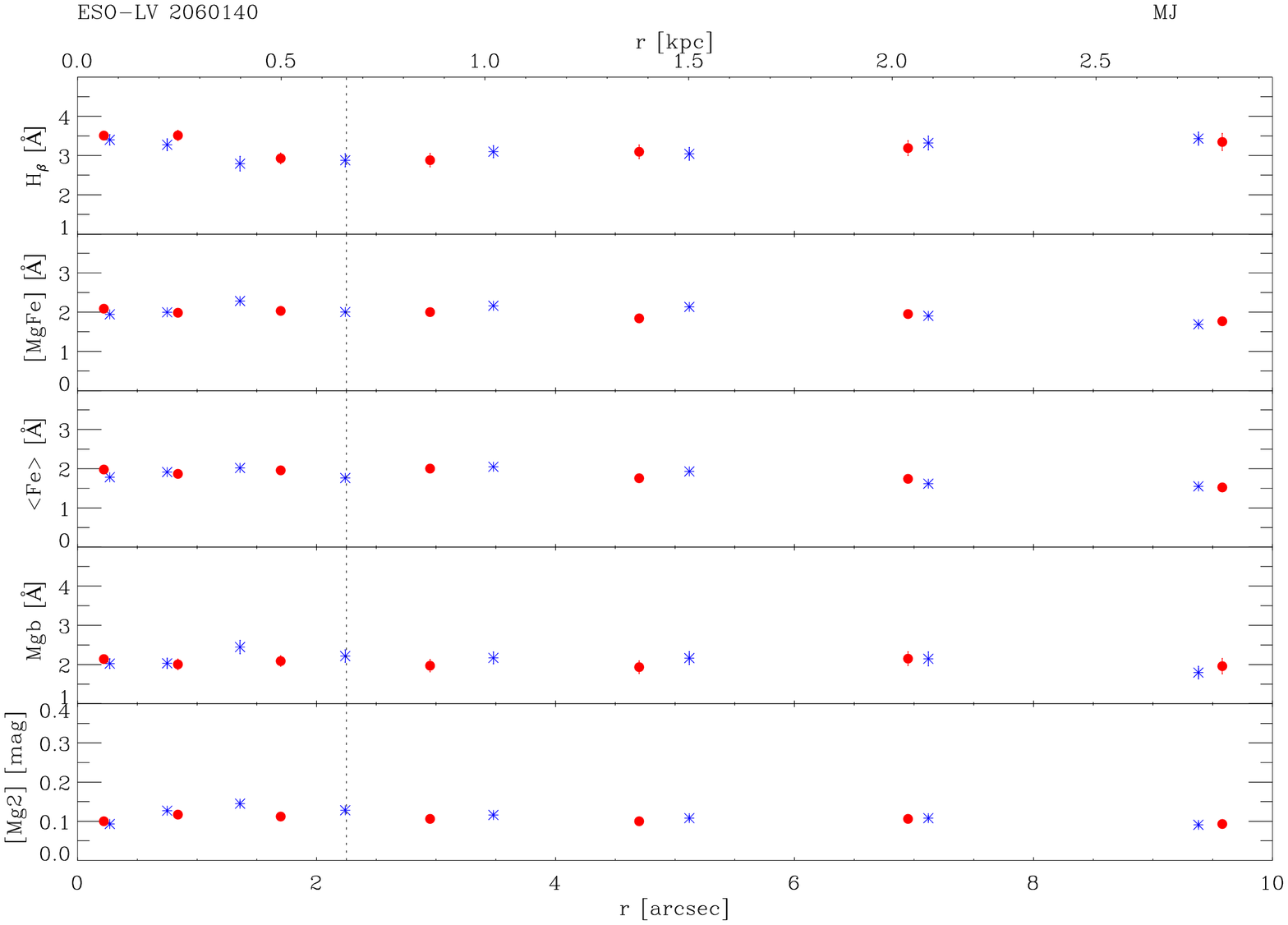}
\includegraphics[angle=90.0,width=0.431\textwidth]{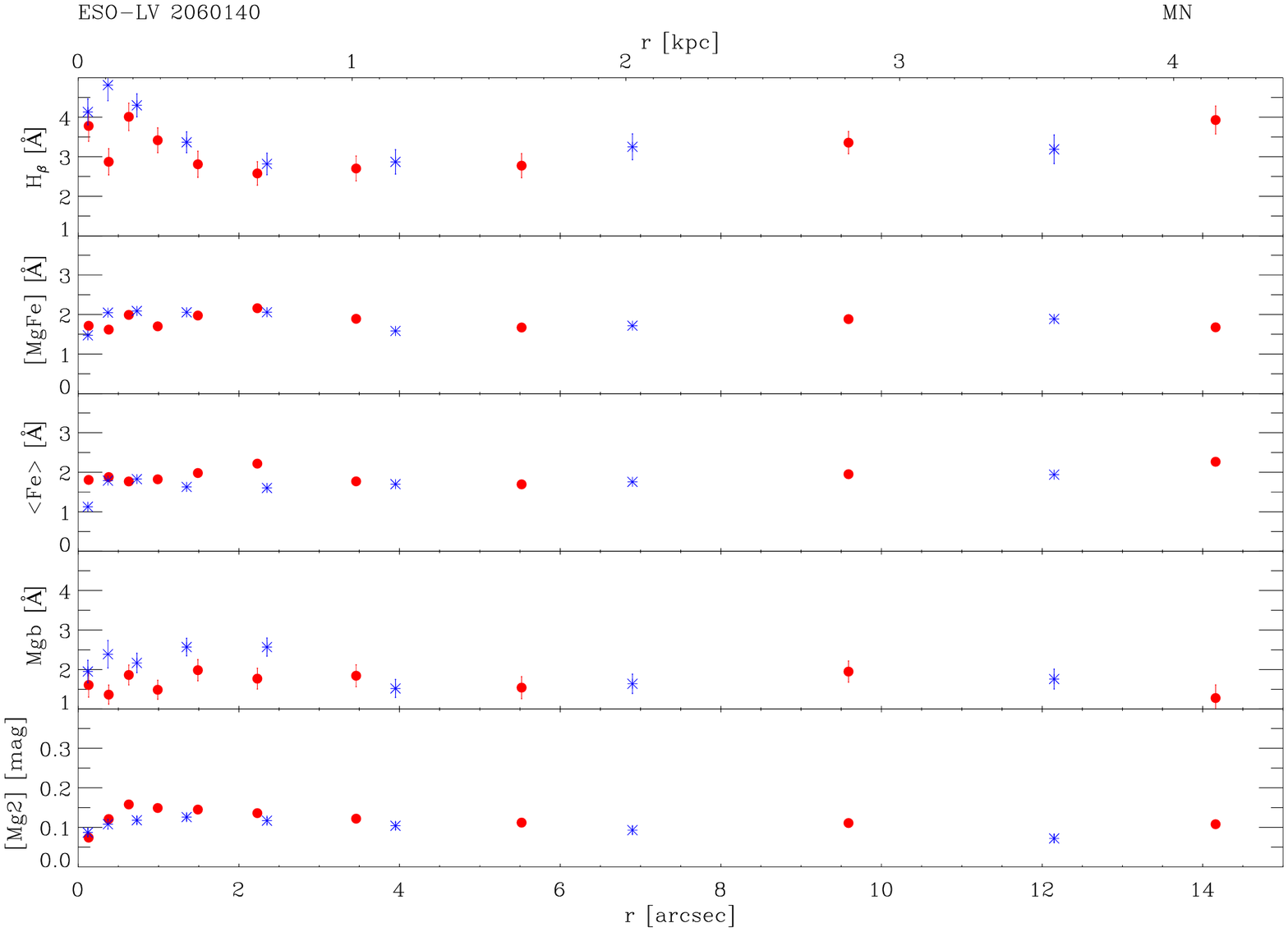}\\
\caption{Line-strength indices measured along the available axes of
  the sample galaxies. For each axis the curves are folded around the
  nucleus. Asterisks and dots refer to data measured along the
  approaching and residing side of the galaxy, respectively.  The
  radial profiles of the line-strength indices \Hb, \MgFe, \Fe, \Mgb,
  and \Mgd\ are shown (from top to bottom panel).  The vertical dashed
  line corresponds to the radius \Rbd\ where the surface-brightness
  contributions of the bulge and disc are equal. For each dataset the
  name of the galaxy and the location of the slit position (Mj=major axis, MN=minor axis) are given.}
\label{fig:indices}
\end{figure*}

\begin{figure*}
\centering
\includegraphics[angle=90.0,width=0.431\textwidth]{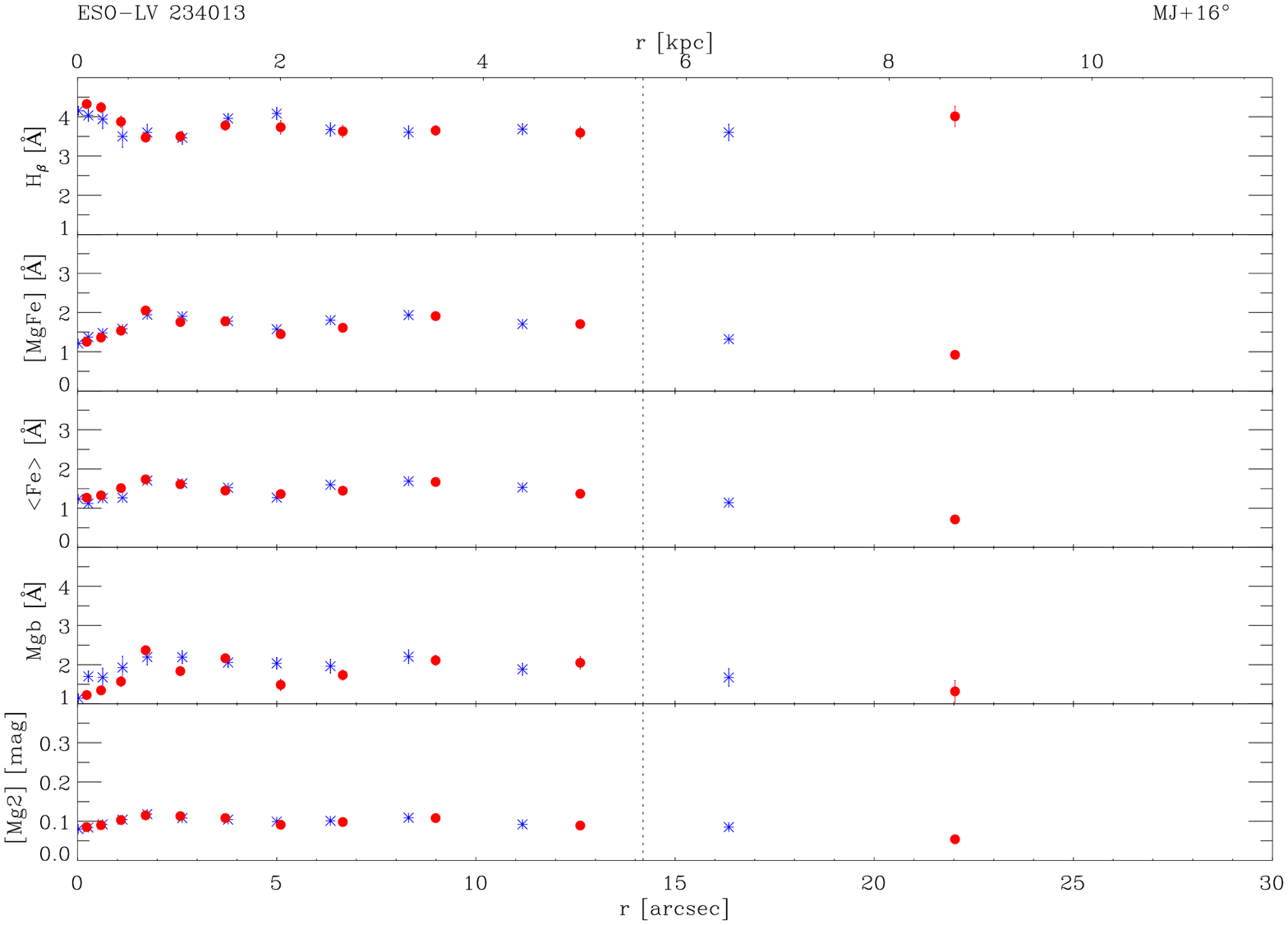}
\includegraphics[angle=90.0,width=0.431\textwidth]{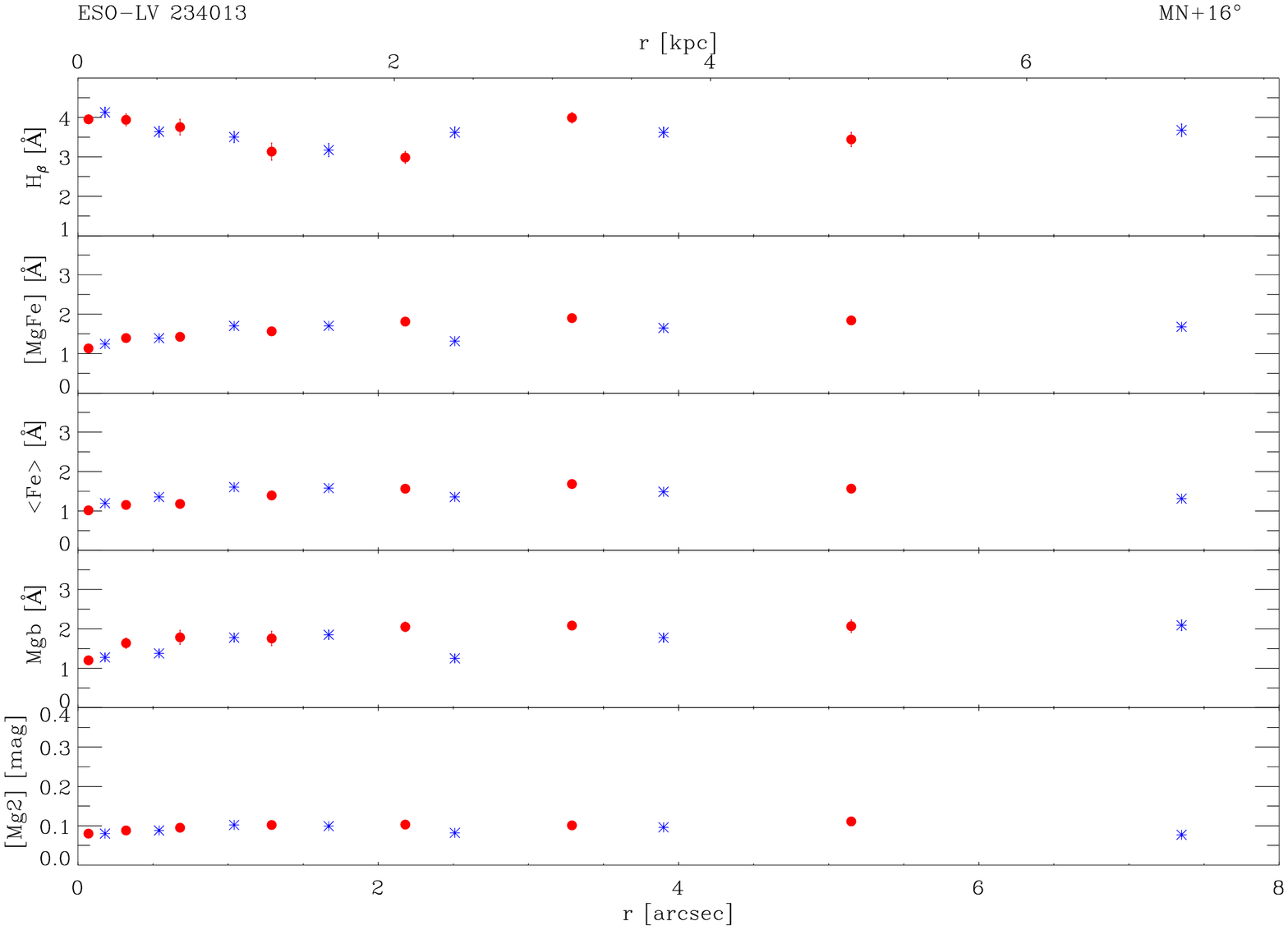}
\includegraphics[angle=90.0,width=0.431\textwidth]{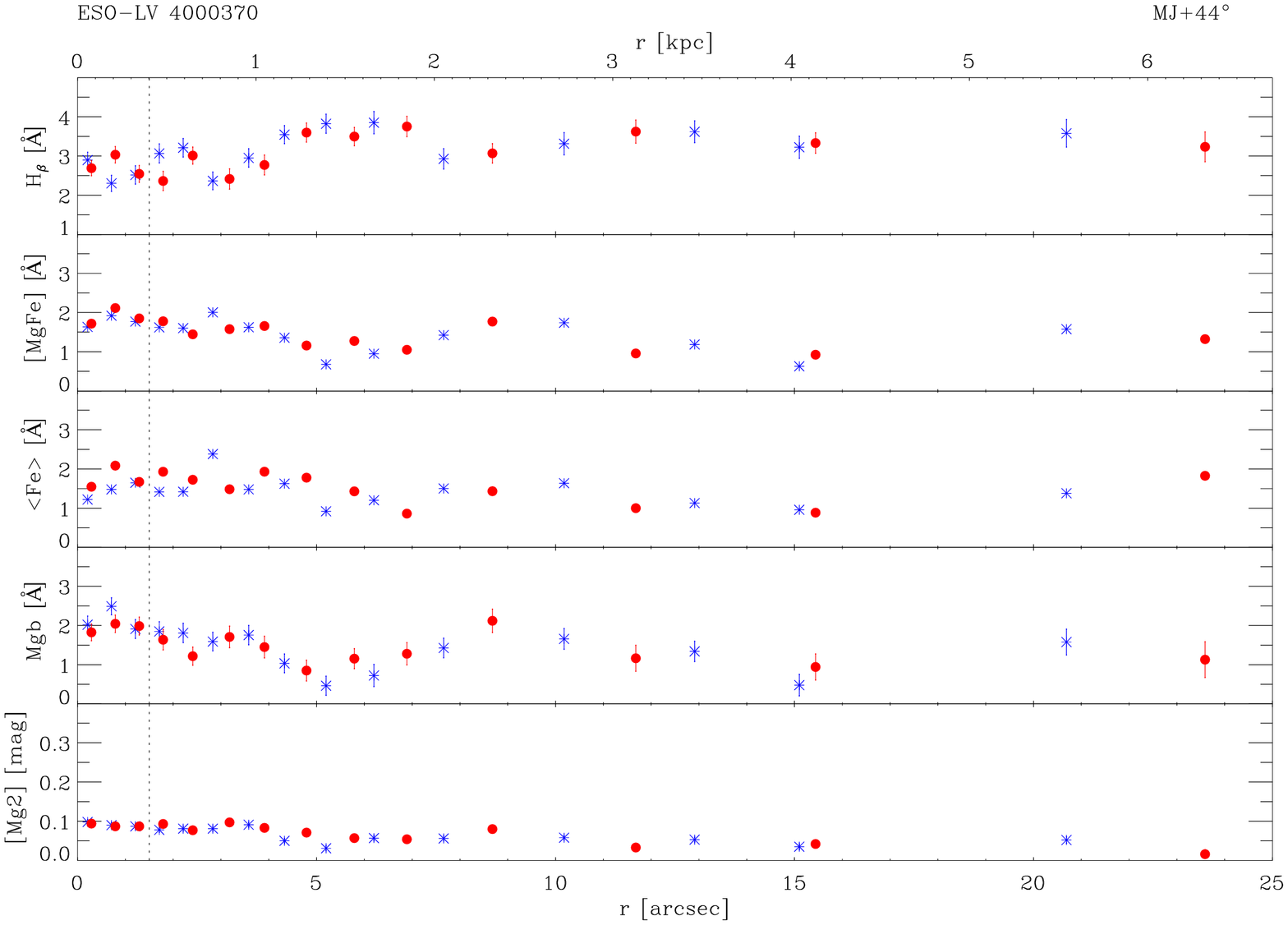}
\includegraphics[angle=90.0,width=0.431\textwidth]{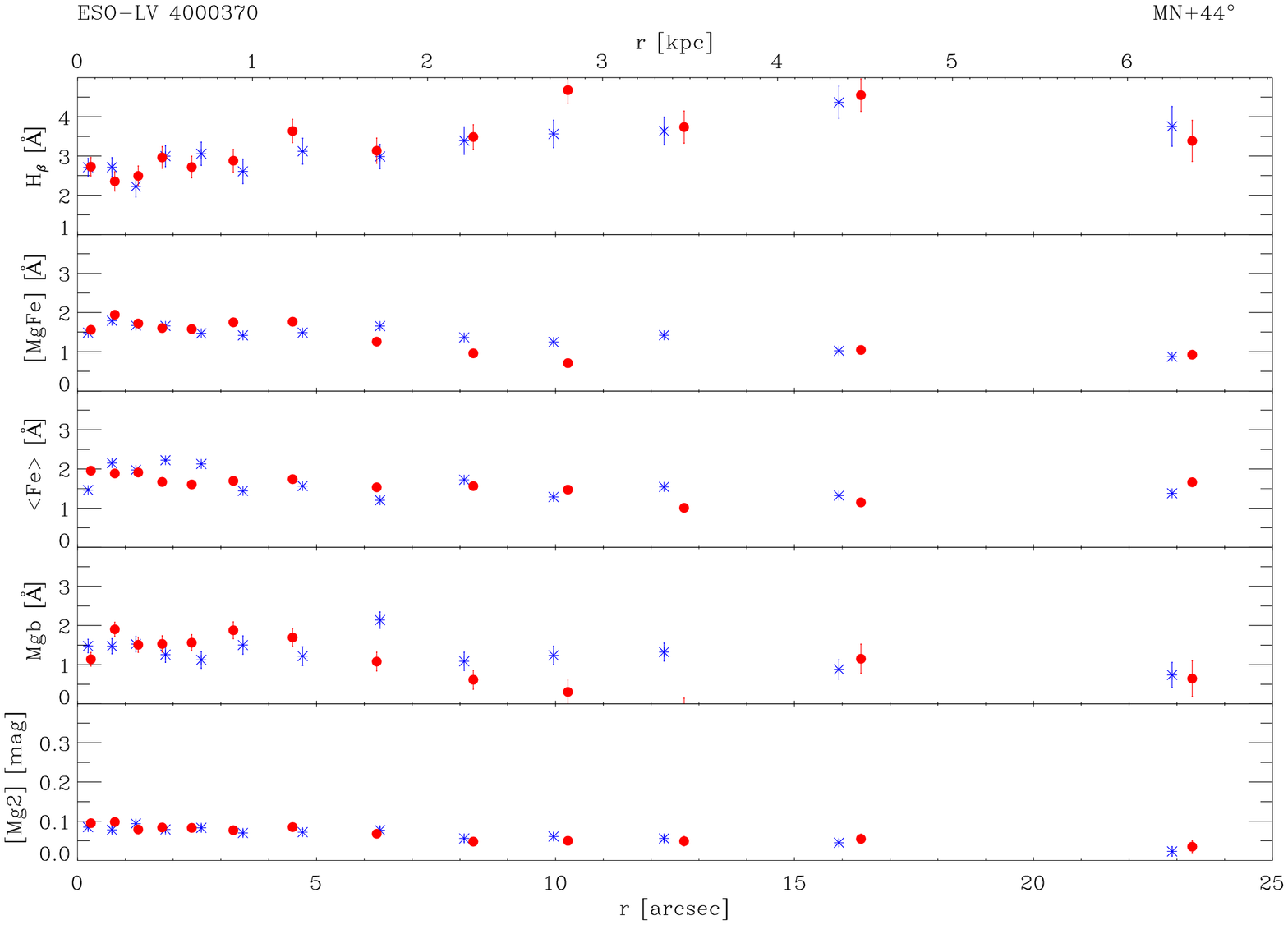}
\includegraphics[angle=90.0,width=0.431\textwidth]{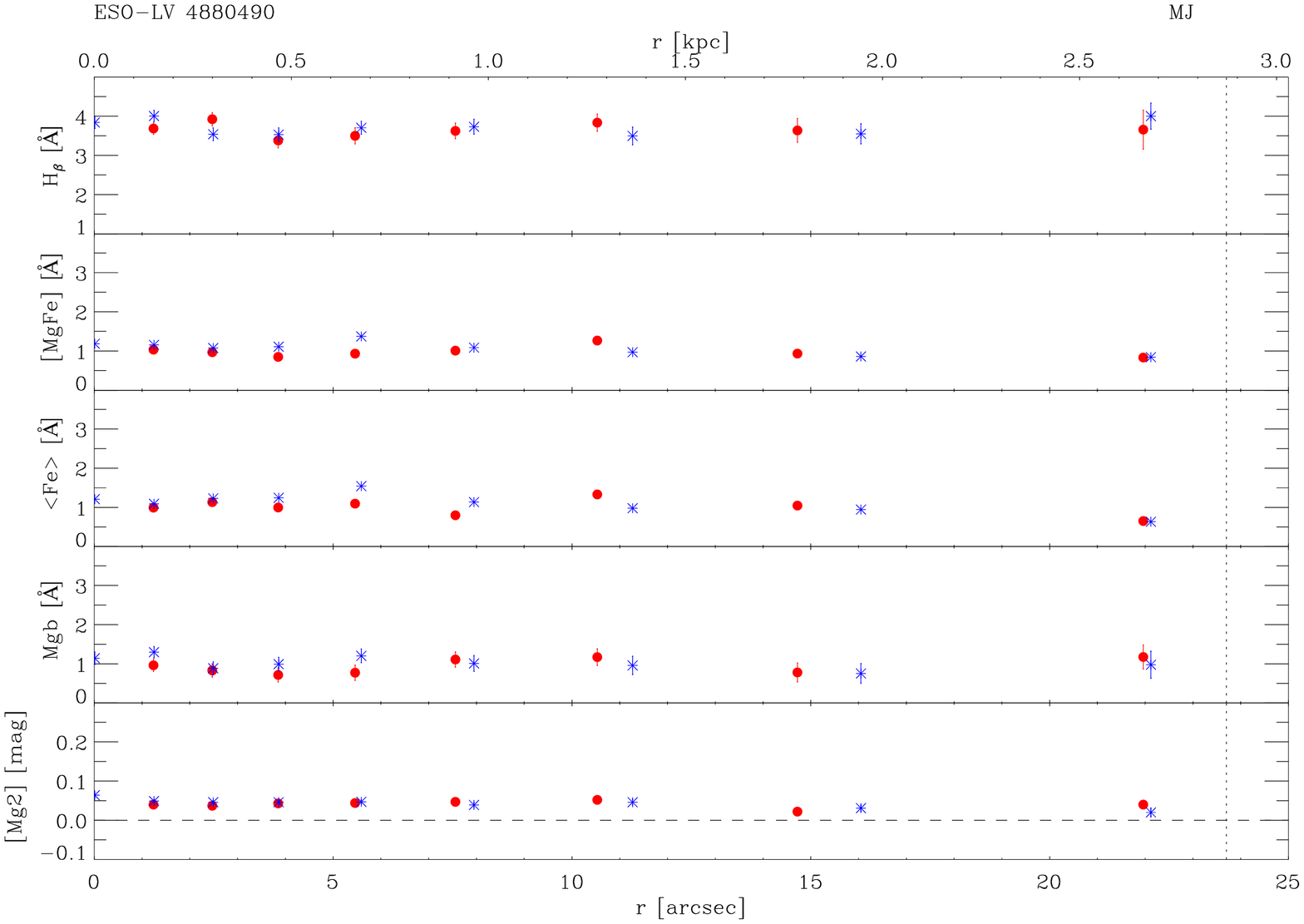}
\includegraphics[angle=90.0,width=0.431\textwidth]{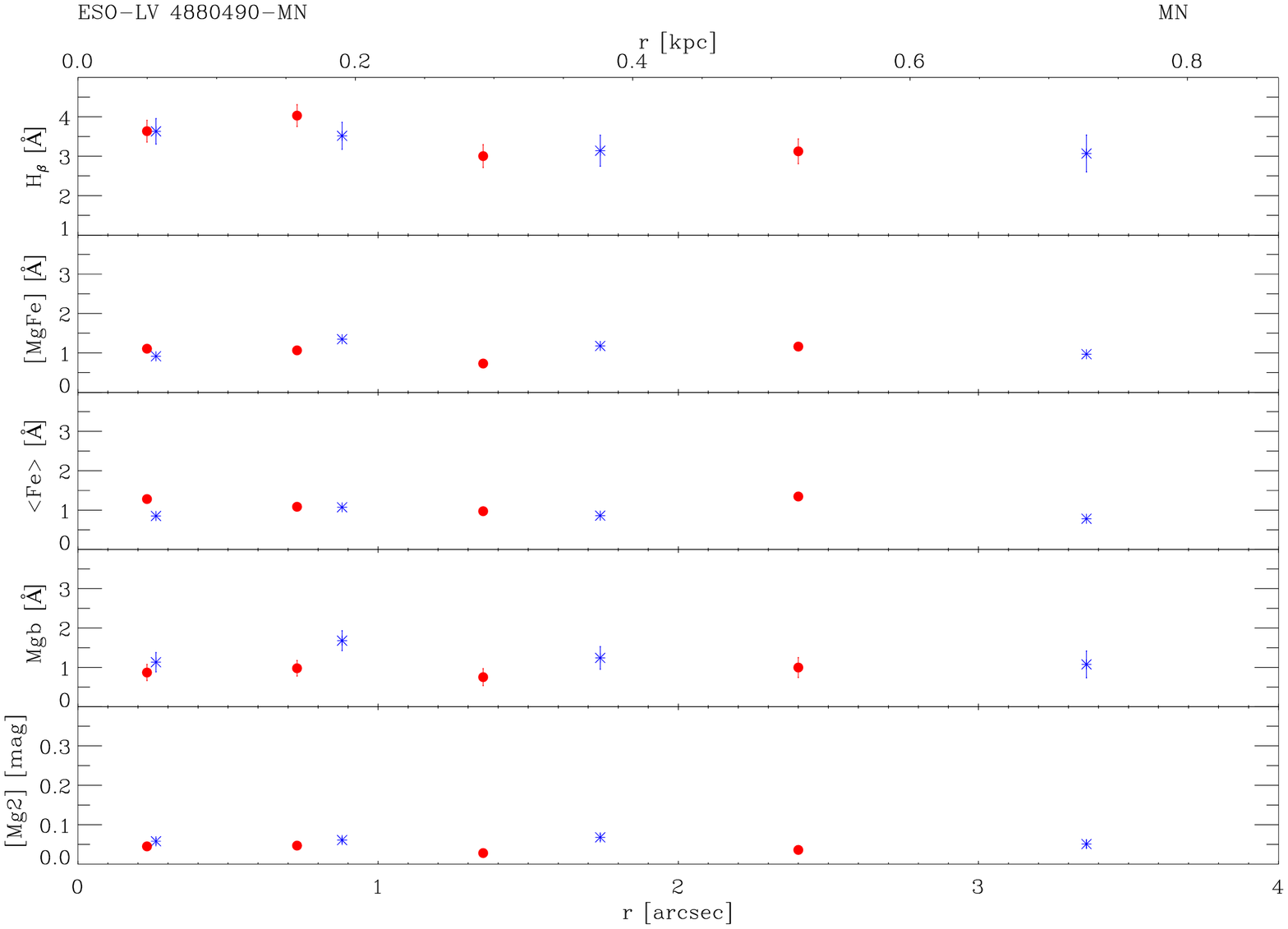}
\includegraphics[angle=90.0,width=0.431\textwidth]{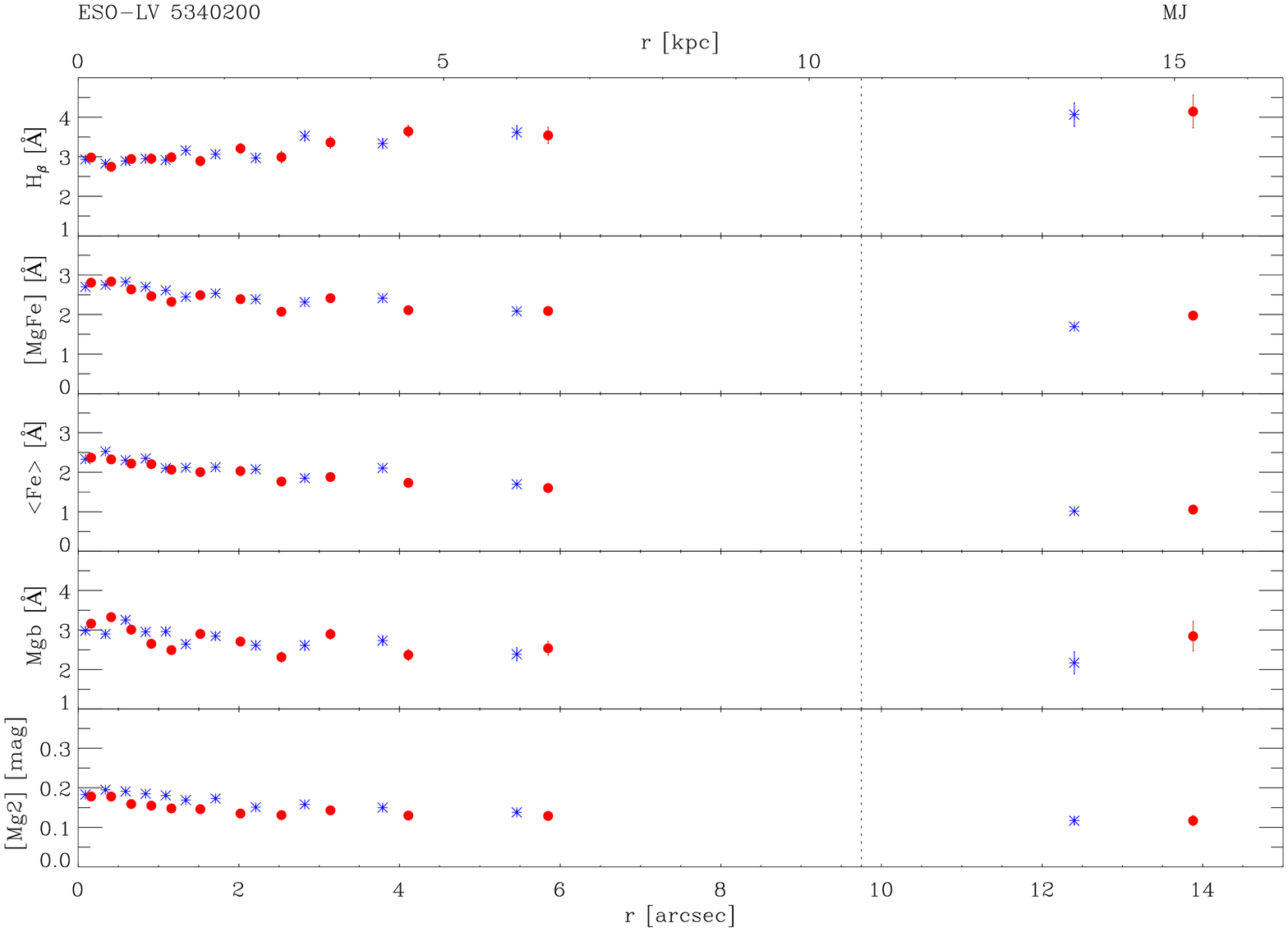}
\includegraphics[angle=90.0,width=0.431\textwidth]{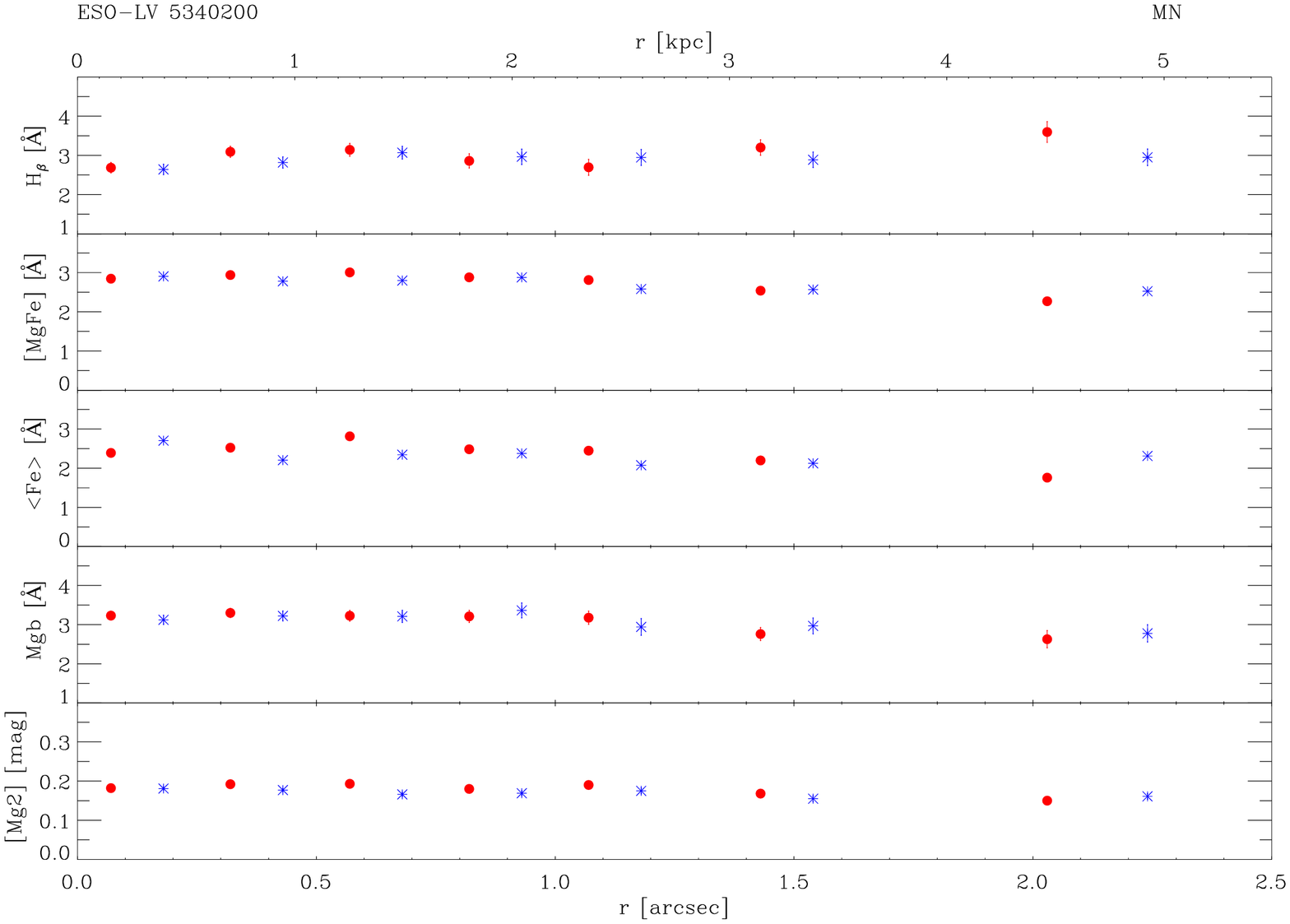}\\
\contcaption{}
\end{figure*}

%%%%%%%%%%%%%%%%%%%%%%%%%%%%%%%%%%%%%%%%%%%%%%%%%%%%%%%%%%%%%%%%%%%%%%%%%%%%%%%%

\section{Properties of the stellar populations}
\label{sec:stellarpop}

\subsection{Central values of the line-strength indices}
\label{sec:linestreng_cent}

In Paper I the central values of velocity dispersion was obtained by a weighted
mean of all the measured data points within an aperture of radius
$0.3\,r_{\rm e}$ along all the available spectra. The adopted weight
for each data point was proportional to the $S/N$ of the spectrum
extracted along the spatial direction where the kinematics measurement
was performed. Similarly, the central value of $\sigma$ for PGC~26148
and PGC~37759, and of \Mgb, \Mgd, \Hb, \Fe, and \MgFe\/ for all the
sample galaxies were obtained in this paper. They are listed in
Tab. \ref{tab:centval_lickind}.

\begin{figure}
\centering
\includegraphics[angle=90,width=0.44\textwidth]{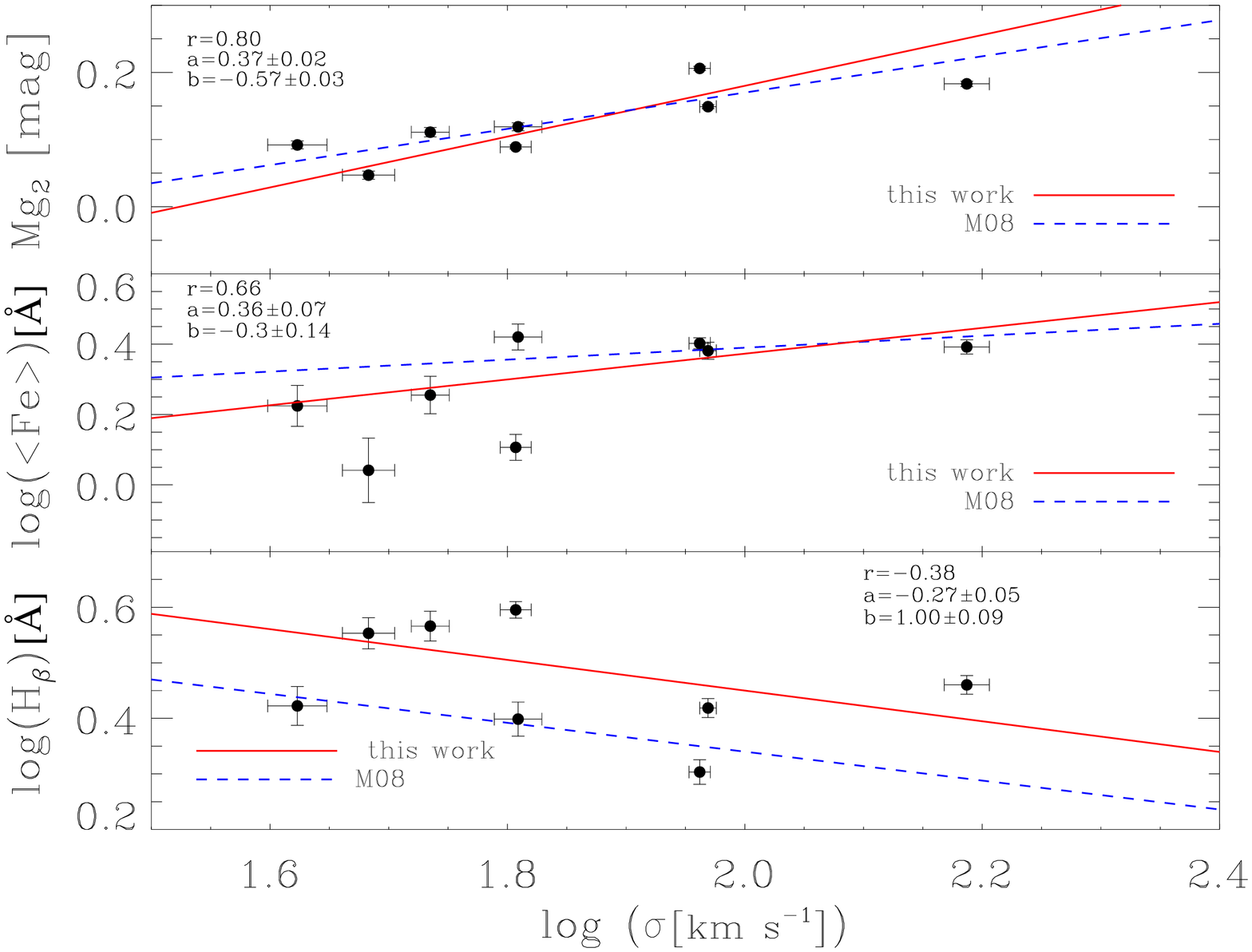}
\caption[The central value correlation \Mgd-$\sigma$ and
  \Fe-$\sigma$]{Central values of the line-strength indices
  \Mgd\ (upper panel), \Fe\ (central panel), and \Hb\ (lower panel) as
  a function of the central velocity dispersion. In each panel the red
  solid line and blue dashed line represent the linear regression
  ($y=ax+b$) through the data points and the correlation found by
  \citet{moreetal08} for bulges of HSB galaxies, respectively. The
  Pearson correlation coefficient ($r$) and the coefficients of the
  linear fit are given.
\label{fig:censmg2fehb}}
\end{figure}

Fig. \ref{fig:censmg2fehb} shows the correlations between the central
values of velocity dispersion and those of \Mgd , \Hb, and \Fe . They are
compared with the results obtained by \citet{moreetal08} for a sample
of bulges hosted by HSB discs. 
Usually, \Mgd\ is adopted as the tracer of the $\alpha$ elements and
gives an estimate of the \aFe\ enhancement, while $\sigma$ is related
to the gravitational potential and, therefore, is a proxy of the mass.
For early-type galaxies and bulges the \Mgd$-\sigma$ correlation
shows that more massive systems host a more metal-rich stellar
population \citep[see][]{idiaetal96, bernetal98, jorgen99, mehletal03,
  moreetal08}. We find that the bulges of LSB galaxies follow a
\Mgd$-\sigma$ relationship which is remarkably similar to that of
their HSB counterparts \citep[e.g.][]{gandetal07, moreetal08}.
Theoretical models of galaxy formation through dissipative collapse
predict a tight \Fe$-\sigma$ correlation \citep[e.g.,][]{kodaetal98},
which is observed for bulges of spiral galaxies \citep{idiaetal96,
  prugetal01, procetal02, moreetal08} but not for early-type galaxies
\citep{fishetal96, jorgen99, tragetal98, mehletal03}. The slope and
zero-point of the \Fe$-\sigma$ relation of the bulges of LSB and in HSB
galaxies \citep{moreetal08} are consistent within errors.
We measure an anti-correlation between \Hb\ and $\sigma$
for the bulges in LSB galaxies. It suggests that less massive bulges
host younger stellar populations with respect to more massive ones. In
spite of the large scatter of the data points, the measured trend is in
good agreement with the measurements obtained by \citet{gandetal07}
and \citet{moreetal08} for bulges in HSB galaxies.

%%%%%%%%%%%%%%%%%%%%%%%%%%%%%%%%%%%%%%%%%%%%%%%%%%%%%%%%%%%%%%%%%%%%%%%%%%%%%%%%
%%%%%%%%%%%%%%%%%%%%%%%%%%%%%%%%%%%%%%%%%%%%%%%%%%%%%%%%%%%%%%%%%%

%%%%%%%%%%%%%%%%%%%%%%%%%%%%%%%%%%%%%%%%%%%%%%%%%%%%%%%%%%%%%%%%%%%%%%%%%%%%%%%%
%%TABELLA VALORI CENTRALI
\begin{table*}
\caption{Central values of the velocity dispersion and line-strength
  indices of the sample galaxies measured within an aperture of radius
  $0.3\,r_{\rm e}$. Column (8) gives the source of the spectrum
  measured in this paper and the reference for $r_{\rm e}$ and
  $\sigma$: 1 = Paper I, 2 = this paper.}
\begin{tabular}{lrcccccc}
\hline
\noalign{\smallskip}
\multicolumn{1}{c}{Galaxy} &
\multicolumn{1}{c}{$\sigma$} &
\multicolumn{1}{c}{\Fe} &
\multicolumn{1}{c}{\MgFe} &
\multicolumn{1}{c}{\Mgd} &
\multicolumn{1}{c}{\Mgb} &
\multicolumn{1}{c}{\Hb} &
\multicolumn{1}{c}{Source} \\
\multicolumn{1}{c}{ } &
\multicolumn{1}{c}{(\kms)} &
\multicolumn{1}{c}{(\AA)} &
\multicolumn{1}{c}{(\AA)} &
\multicolumn{1}{c}{(mag)} &
\multicolumn{1}{c}{(\AA)} &
\multicolumn{1}{c}{(\AA)} &
\multicolumn{1}{c}{} \\
\multicolumn{1}{c}{(1)} &
\multicolumn{1}{c}{(2)} &
\multicolumn{1}{c}{(3)} &
\multicolumn{1}{c}{(4)} &
\multicolumn{1}{c}{(5)} &
\multicolumn{1}{c}{(6)} &
\multicolumn{1}{c}{(7)} &
\multicolumn{1}{c}{(8)} \\
\noalign{\smallskip}
\hline
\noalign{\smallskip}
ESO-LV~1860550 & $ 91.7\pm2.0$ & $2.53\pm0.09$ & $2.97\pm0.01$ & $0.206\pm0.003$ & $3.40\pm0.09$ & $2.01\pm0.10$ & 1\\
ESO-LV~2060140 & $ 54.3\pm2.0$ & $1.80\pm0.23$ & $1.92\pm0.05$ & $0.111\pm0.007$ & $1.96\pm0.20$ & $3.68\pm0.23$ & 1\\
ESO-LV~2340130 & $ 64.1\pm2.0$ & $1.28\pm0.11$ & $1.39\pm0.01$ & $0.089\pm0.003$ & $1.49\pm0.11$ & $3.94\pm0.14$ & 1\\
ESO-LV~4000370 & $ 42.0\pm2.5$ & $1.68\pm0.24$ & $1.76\pm0.04$ & $0.092\pm0.006$ & $1.78\pm0.20$ & $2.64\pm0.22$ & 1\\
ESO-LV~4880490 & $ 48.2\pm2.5$ & $1.10\pm0.26$ & $1.06\pm0.03$ & $0.047\pm0.006$ & $1.02\pm0.19$ & $3.57\pm0.24$ & 1\\
ESO-LV~5340200 & $153.9\pm7.0$ & $2.47\pm0.12$ & $2.80\pm0.02$ & $0.183\pm0.004$ & $3.14\pm0.10$ & $2.86\pm0.11$ & 1\\
PGC~26148      & $ 93.1\pm1.6$ & $2.41\pm0.14$ & $2.59\pm0.02$ & $0.149\pm0.003$ & $2.73\pm0.11$ & $2.62\pm0.10$ & 2\\
PGC~37759      & $ 64.4\pm3.1$ & $2.63\pm0.23$ & $2.97\pm0.07$ & $0.119\pm0.006$ & $3.17\pm0.20$ & $2.50\pm0.18$ & 2\\
\noalign{\smallskip}
\hline
\noalign{\bigskip}
\label{tab:centval_lickind}
\end{tabular}
\end{table*}
%%%%%%%%%%%%%%%%%%%%%%%%%%%%%%%%%%%%%%%%%%%%%%%%%%%%%%%%%%%%%%%%%%%%%%%%%%%%%%%%

\subsection{Central values of the age, metallicity, and ${\bf \alpha}$/Fe enhancement} 
\label{agemet_cent}

\begin{figure}
\centering
\includegraphics[angle=90.0,width=0.495\textwidth]{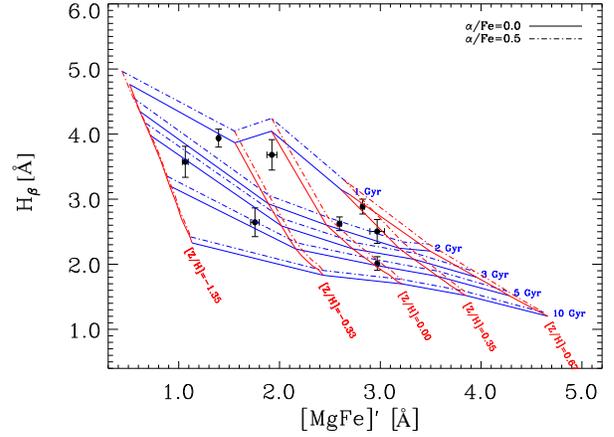}
\includegraphics[angle=90.0,width=0.495\textwidth]{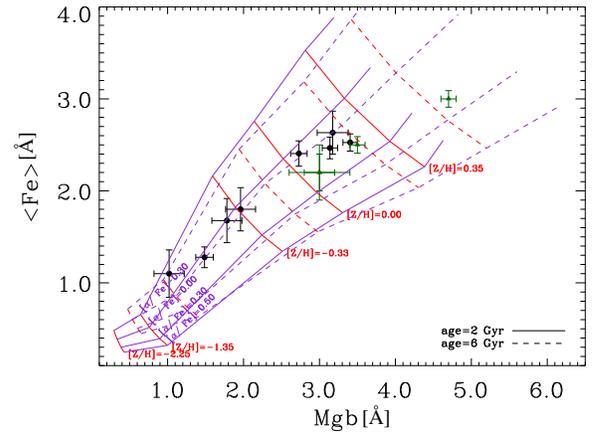}
\caption[Distribution \Hb-\MgFe\/ and \Mgb-\Fe]{The distribution of
  the central values of \Hb\/ and \MgFe\/ indices (top panel) and \Fe
  and \Mgb\/ indices (bottom panel) measured over an aperture of
  $0.3\,r_{\rm e}$ for the sample galaxies. The lines indicate the
  models by \cite{thmabe03}. In the top panel the age-metallicity
  grids are plotted with two different $\alpha$/Fe enhancements:
  \aFe$\,=\,0.0$ dex (continuous lines) and \aFe$\,=\,0.5$ dex (dashed
  lines). In the bottom panel the \aFe\ ratio-metallicity grids are
  plotted with two different ages: 2 Gyr (continuous lines) and 6 Gyr
  (dashed lines). Green triangles are the values obtained for 3 LSB
  galaxies by \citet{bergetal03}. 
\label{fig:hbemgfemgbcent}}
\end{figure}

\begin{figure}%[hp!]
\centering
\includegraphics[angle=90,width=0.48\textwidth]{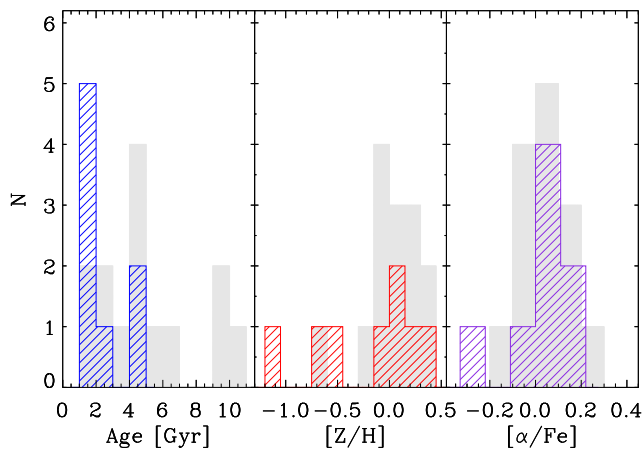}\\
\caption[]{Distribution of the mean age (left-hand panel), total
  metallicity (central panel), and total $\alpha$/Fe enhancement
  (right-hand panel) for the stellar populations of the bulges of the
  sample galaxies. The distribution of the same
  quantities obtained by \citet{moreetal08} for the bulges of HSB
  discs are plotted in grey for sake of comparison.
\label{fig:hist_ama}}
\end{figure}

The models by \citet{thmabe03} predict the values of the line-strength
indices for a single stellar population as function of the age,
metallicity, and \aFe\ ratio. 
In the top panel of Fig. \ref{fig:hbemgfemgbcent} the central values
of \Hb\ and \MgFe\ are compared with the model predictions by
\citet{thmabe03} for two stellar populations with solar (\aFe$\,=\,0$
dex) and super-solar $\alpha/$Fe enhancement (\aFe$\,=\,0.5$ dex),
respectively. In this parameter space the mean age and total
metallicity appear to be almost insensitive to the variations of the
$\alpha/$Fe enhancement.
In the bottom panel of Fig. \ref{fig:hbemgfemgbcent} the central
values of \Mgb\ and \Fe\ are compared with the model predictions by
\citet{thmabe03} for two stellar populations with an intermediate (2
Gyr) and old age (6 Gyr), respectively. In this parameter space the
total metallicity and total $\alpha/$Fe enhancement appear to be
almost insensitive to the variations of the age.
The central mean age, total metallicity, and total $\alpha/$Fe
enhancement of the stellar population in the centre of the sample
galaxies were derived from the values of line-strength indices given
in Table \ref{tab:centval_lickind} by a linear interpolation between
the model points using the iterative procedure described in
\citet{mehletal03} and \citet{moreetal08}. The derived values and
their corresponding errors are representative of the properties of the
stellar populations of the galaxy bulges and listed in
Tab. \ref{tab:agemetalfa}. The histograms of their number distribution
are plotted in Fig. \ref{fig:hist_ama}.

%%%%%%%%%%%%%%%%%%%%%%%%%%%%%%%%%%%%%%%%%%%%%%%%%%%%%%%%%%%%%%%%%%%%%%%%%
%%% TABLE Indeces central values
%%%%%%%%%%%%%%%%%%%%%%%%%%%%%%%%%%%%%%%%%%%%%%%%%%%%%%%%%%%%%%%%%

\begin{table}
\caption{Mean age, total metallicity, and total $\alpha/$Fe
  enhancement of the stellar populations of the bulges of the sample
  galaxies}
\begin{center}
\begin{small}
\begin{tabular}{lrrr}
\hline
\noalign{\smallskip}
\multicolumn{1}{c}{Galaxy} &
\multicolumn{1}{c}{Age} &
\multicolumn{1}{c}{\ZH} &
\multicolumn{1}{c}{\aFe} \\
\noalign{\smallskip}
\multicolumn{1}{c}{} &
\multicolumn{1}{c}{[Gyr]} &
\multicolumn{1}{c}{} &
\multicolumn{1}{c}{} \\
\noalign{\smallskip}
\multicolumn{1}{c}{(1)} &
\multicolumn{1}{c}{(2)} &
\multicolumn{1}{c}{(3)} &
\multicolumn{1}{c}{(4)} \\
\noalign{\smallskip}
\hline
\noalign{\smallskip}  
ESO-LV~1860550 &$4.9\pm1.6$ & $ 0.01\pm0.06$  & $ 0.11\pm0.06$ \\
ESO-LV~2060140 &$1.3\pm0.2$ & $-0.11\pm0.07$  & $ 0.06\pm0.13$ \\
ESO-LV~2340130 &$1.2\pm0.2$ & $-0.53\pm0.06$  & $ 0.09\pm0.09$ \\
ESO-LV~4000370 &$4.3\pm1.4$ & $-0.62\pm0.09$  & $-0.04\pm0.12$ \\
ESO-LV~4880490 &$2.9\pm1.0$ & $-1.07\pm0.09$  & $-0.24\pm0.20$ \\
ESO-LV~5340200 &$1.5\pm0.1$ & $ 0.31\pm0.04$  & $ 0.18\pm0.06$ \\
PGC~26148      &$1.9\pm0.2$ & $ 0.08\pm0.04$  & $ 0.05\pm0.06$ \\
PGC~37759      &$1.8\pm0.6$ & $ 0.29\pm0.11$  & $ 0.10\pm0.11$ \\
\noalign{\smallskip}
\hline
\noalign{\medskip}
\end{tabular}
\end{small}
\label{tab:agemetalfa}
\end{center}
\end{table}
%%%%%%%%%%%%%%%%%%%%%%%%%%%%%%%%%%%%%%%%%%%%%%%%%%%%%%%%%%%%
%%%%%%%%%%%%%%%%%%%%%%%%%%%%%%%%%%%%%%%%%%%%%%%%%%%%%%%%%%%%

All the bulges are characterized by a very young stellar population,
with a distribution of ages peaked at the value of 1.5 Gyr. Only the
bulges of ESO-LV~1860550 and ESO-LV~4000370 show an intermediate-age
(4-5 Gyr) stellar population.  The globally young nature of these
objects is also suggested from the ongoing star formation as shown by
the presence of the \Hb\ emission line in their spectra. This result
confirms the presence of young stars in LSB galaxies as previously
reported from spectroscopic \citep{bergetal03, zhonetal10} and
photometric \citep{galaetal02,
  zacketal05,galaetal06,voroetal09,galaetal11} analysis of the stellar
populations. Even though the number of sample galaxies does not allow
us to trace a firm statistical conclusion, it is interesting to note
that we do not find any old-age bulge in LSB discs
(Fig. \ref{fig:hist_ama}, left-hand panel), whereas they were
approximately 25 per cent in the HSB sample studied with the same
technique by \citet{moreetal08} which is similar to our sample for
both mass range and morphological type.  The metallicity of the bulges
of LSB discs spans a large range of values from high (\ZH$\,=\,0.30$
dex) to sub-solar metallicity (\ZH$\,=-1.0$ dex) with an almost flat
distribution (Fig. \ref{fig:hist_ama}, middle panel).  The obtained
ages and metallicity well match the $B-r$ colour (or $B-R$ given in
Paper I) derived for the sample galaxies.

Most of the sample bulges display solar $\alpha/$Fe enhancements
(Fig. \ref{fig:hist_ama}, right-hand panel). The number distribution
has a median at \aFe$\;=0.09$ and spreads from super (\aFe$\,=\,0.3$)
to sub-solar values (\aFe$\,=\,-0.2$). These values are consistent
with those obtained for the 3 LSB galaxies studied by
\citet{bergetal03} and plotted for comparison in
Fig. \ref{fig:hbemgfemgbcent}. They are also remarkably similar to the
\aFe\ ratios derived for the bulges of HSB galaxies \citep{moreetal08}
and for elliptical galaxies in clusters \citep{tesipeletier, jorg99,
  tragetal00, kuntetal00, kuntetal02}. They imply a star-formation
time-scale ranging from less than 1 to 5 Gyr in agreement with the
predictions of most of the models of bulge formation \citep{gilwys98,
  elmetal08, cesmat11}.

%%%%%%%%%%%%%%%%%%%%%%%%%%%%%%%%%%%%%%%%%%%%%%%%%%%%%%%%%%%%
% Figure histogramma valori eta metallicita alfa enhancement
%%%%%%%%%%%%%%%%%%%%%%%%%%%%%%%%%%%%%%%%%%%%%%%%%%%%%%%%%%%%

\begin{figure}%[hp!]
\centering
\includegraphics[angle=90,width=0.5\textwidth]{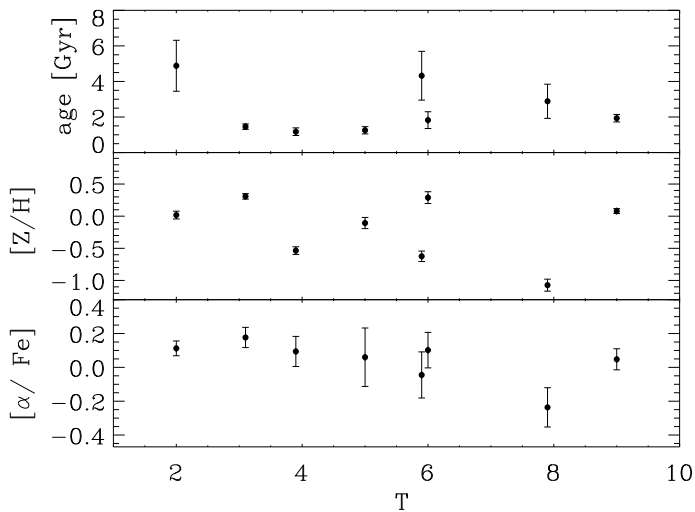}\\
\caption[]{Mean age (upper panel), total metallicity (middle panel),
  and total $\alpha/$Fe enhancement (lower panel) of the stellar
  populations of the bulges of the sample galaxies as a function of galaxy
  morphological type. 
\label{fig:T_agemetalfa}}
\end{figure}

The correlation between the galaxy morphological type and the
properties of the bulge stellar population is indicative of the
possible interplay between the evolution of bulge and
disc. \citet{thda06} did not observe any correlation between the age
and metallicity of the bulge stellar population and galaxy morphology,
whereas \citet{gandetal07} and \citet{moreetal08} found a mild
correlation with the early-type galaxies ($T\,<\,0$) being older and
more metal rich than spiral galaxies ($T\,\geq\,0$). Since the above
relationships are mostly driven by the early-type galaxies which are
lacking in our sample, we do not observe any correlation between the
galaxy morphological type ($2\,\leq\,T\,\leq\,9$) and the age,
metallicity, or $\alpha/$Fe enhancement of our bulges
(Fig. \ref{fig:T_agemetalfa}). Nevertheless, we can exclude a strong
interplay between the bulge and disc components.

%%%%%%%%%%%%%%%%%%%%%%%%%%%%%%%%%%%%%%%%%%%%%%%%%%%%%%%%%%%%
%figure di T contro eta metallicita alfa
%%%%%%%%%%%%%%%%%%%%%%%%%%%%%%%%%%%%%%%%%%%%%%%%%%%%%%%%%%%%

%%%%%%%%%%%%%%%%%%%%%%%%%%%%%%%%%%%%%%%%%%%%%%%%%%%%%%%%%%%%%%%%%%%%%%%%%
% --- Figure eta metallicita alfe all VS all
\begin{figure*}%[hp!]
\centering
\includegraphics[angle=90,width=1\textwidth]{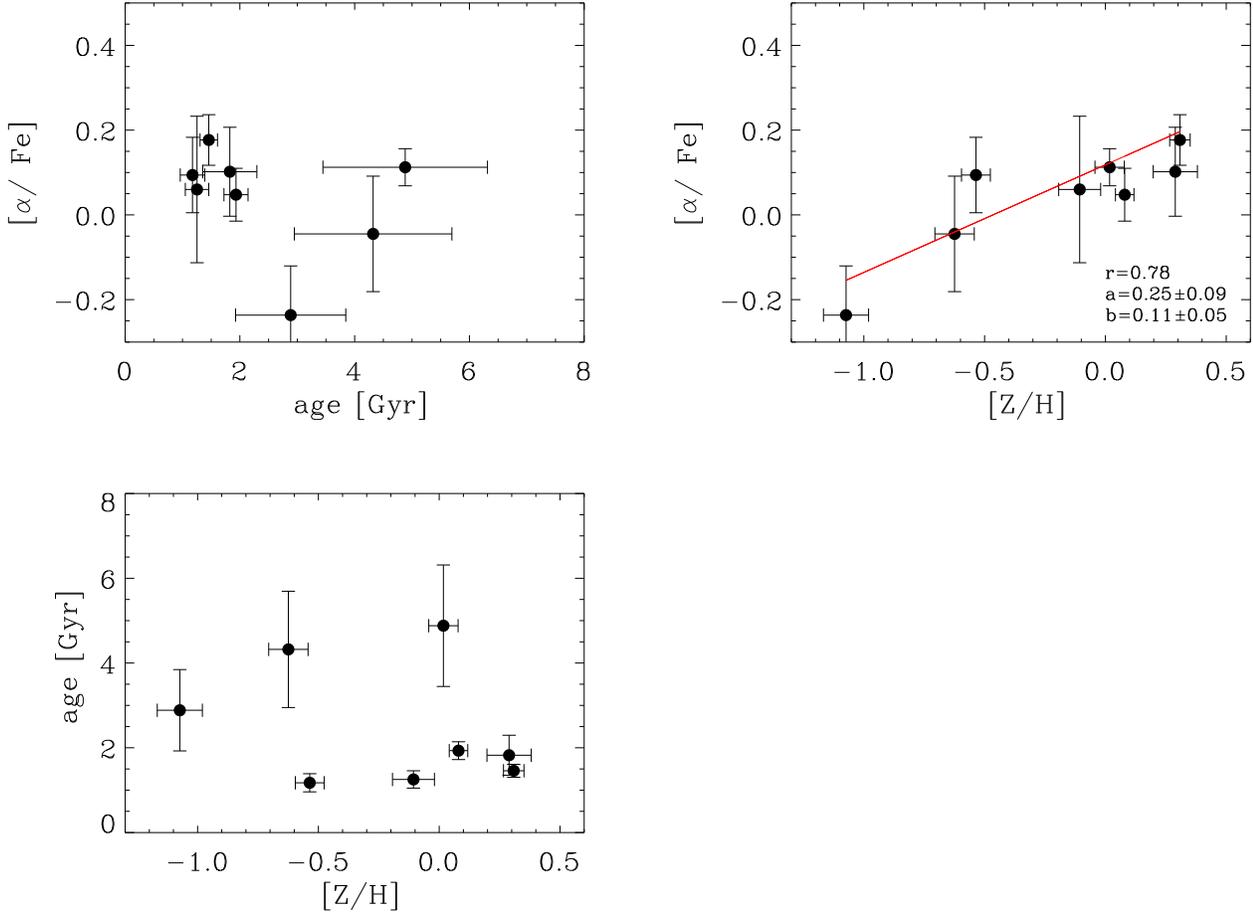}\\
\caption[]{Mean age, total metallicity, and $\alpha/$Fe enhancement of
  the stellar populations of the bulges of the sample galaxies. In the
  upper right-hand panel the red solid line represents the linear
  regression ($y=ax+b$) through the data points. The Pearson
  correlation coefficient ($r$) and the coefficients of the linear fit
  are given.
\label{fig:mgbfe}}
\end{figure*}
%%%%%%%%%%%%%%%%%%%%%%%%%%%%%%%%%%%%%%%%%%%%%%%%%%%%%%%%%%%%%%%%%%%%%%%%%

%%%%%%%%%%%%%%%%%%%%%%%%%%%%%%%%%%%%%%%%%%%%%%%%%%%%%%%%%%%%
%figure di sigma contro eta metallicita alfa
\begin{figure}%[hp!]
\centering
\includegraphics[angle=90,width=0.5\textwidth]{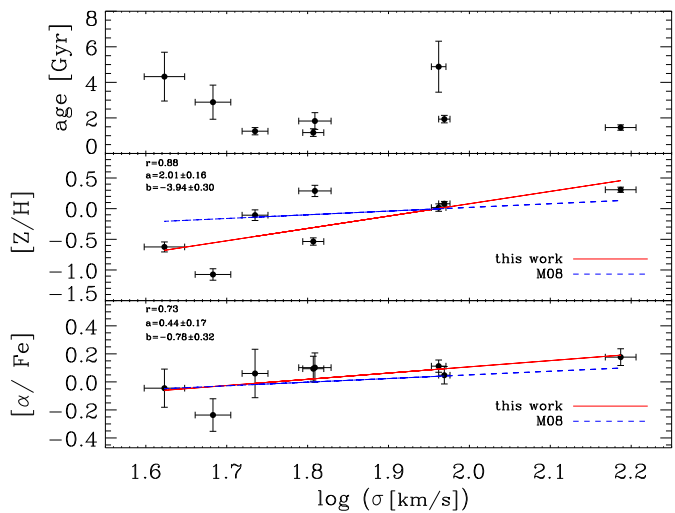}\\
\caption[]{Mean age (upper panel), total metallicity (middle panel),
  and total $\alpha$/Fe enhancement (lower panel) of the stellar
  populations of the bulges of the sample galaxies as a function of the
  central velocity dispersion. In each panel the red solid line and
  blue dashed line represent the linear regression ($y=ax+b$) through
  the data points and the correlation found by \citet{moreetal08} for
  bulges of HSB galaxies, respectively. The Pearson correlation
  coefficient ($r$) and the coefficients of the linear fit are
  given. 
\label{fig:sig_agemetalfa}}
\end{figure}
%%%%%%%%%%%%%%%%%%%%%%%%%%%%%%%%%%%%%%%%%%%%%%%%%%%%%%%%%%%%

Despite their different distributions, metallicity and $\alpha$/Fe
enhancement are strongly correlated, as shown in
Fig. \ref{fig:mgbfe}. This implies that the last episode of star
formation was very short in galaxies with high metallicities, whereas
it lasts longer in galaxies with low metallicity. A similar trend was
also found for the bulges of HSB galaxies \citep{moreetal08} and for
early-type galaxies \citep{spoletal10}. We do not find any correlation
between age and metallicity or $\alpha$/Fe enhancement
(Fig. \ref{fig:mgbfe}).

In early-type galaxies \citep{mehletal03, denietal05, thometal05,
  sancetal06p, annietal07, macaetal09, spoletal10} and in bulges of
HSB galaxies \citep{gandetal07, moreetal08} the metallicity and
$\alpha$/Fe enhancement are well correlated with the central velocity
dispersion. Metallicity and $\alpha/$Fe enhancement of bulges of LSB
galaxies correlate with velocity dispersion
(Fig. \ref{fig:sig_agemetalfa}) and the correlation is consistent with
the results obtained in \citet{moreetal08}. These relations are
explained by chemodinamycal models \citep{matteucci1994, kawgib03,
  kobayashi04} and cosmological hydrodynamic simulations
\citep{deluetal04, tassetal08} of ellipticals and bulges as the result
of a mass-dependent star formation efficiency. Indeed, low-mass
galaxies have a lower efficiency in converting gas-phase metals into
new stars and this gives rise to a prolonged star formation and to
lower $\alpha/$Fe ratios. Our results suggest that this is true also
for the bulges of LSB galaxies. We conclude the most massive bulges of
our sample galaxies are more metal rich, and characterized by a
shorter star-formation time-scale.
The relationship between the age and central velocity dispersion is
more controversial. The presence of an age drop for low-mass galaxies
($\log{\sigma}\,<\,2.1$) was pointed out by \citet{nelaetal05} and was
observed also by \citet{thometal05}, \citet{moreetal08}, and
\citet{spoletal10}. We do not find any correlation, but all our
galaxies are in the low-mass regime, except for ESO-LV~5340200.

The case of bulge of ESO-LV~4880490 is particularly interesting. The
central stellar population of this late-type barred spiral has an
intermediate age (3 Gyr) and very low values of both metal content
(\ZH$\,=\,-1.07$ dex) and $\alpha/$Fe ratio (\aFe$\,=\,-0.24$
dex). This suggests a very prolonged star formation history in the
galaxy centre consistent with a slow building up of the bulge within a
scenario of secular evolution driven by the bar. 

\subsection{Radial gradients of the age, metallicity, and ${\bf \alpha}$/Fe enhancement} 
\label{sec:agemetalpha_grad}

An issue in measuring the gradients of the age, metallicity, and
$\alpha/$Fe enhancement in bulges could be the contamination of their
stellar population by the light coming from the underlying disc
stellar component. This effect is negligible in the galaxy centres but
it could increase going to the outer regions of bulges, where the
light starts to be dominated by the disc component.  In order to
reduce the impact of disc contamination and extend as much as possible
the region in which deriving gradients, we map them inside \Rbd, the
radius where the bulge and disc give the same contribution to the
total surface brightness \citep{moreetal08}. Deriving gradients in the
bulge dominated region with this approach will not remove completely
the contamination by the disc stellar population, but it will assure
always a similar degree of contamination in comparing the gradients of
different galaxies.

\begin{figure}%[hp!]
\centering
\includegraphics[angle=90,width=0.5\textwidth]{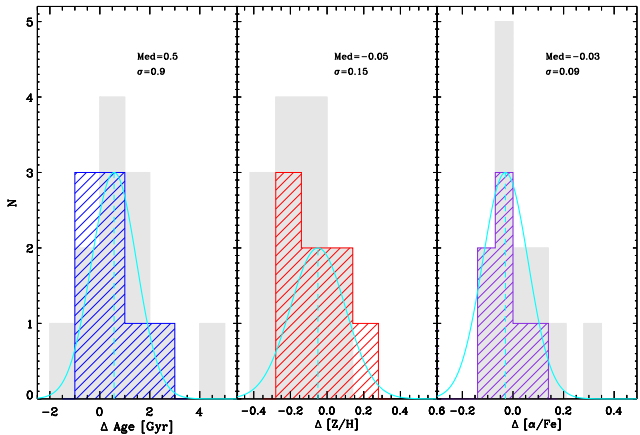}\\
\caption[]{Distribution of the gradients of age (left-hand panel),
  metallicity (central panel) and $\alpha$/Fe enhancement (right-hand
  panel) for the sample bulges. The dashed line represents the median
  of the distribution and its values is reported. Solid line
  represents a Gaussian centred in the median value of
  distribution. Its $\sigma$ approximated by the value containing the
  68\% of the objects of the distribution is reported. In grey is
  represented the distribution of the same quantities obtained by
  \citet{moreetal08} for the bulges of HSB discs. 
\label{fig:histgrad_ama}}
\end{figure}

For each galaxy, we derived the \Mgd , \Hb, and, \Fe\/ line-strength
indices at the radius \Rbd\ (Tab. \ref{tab:agemetalfa_grad}). The
corresponding ages, metallicities, and $\alpha/$Fe enhancements were
derived by using the stellar population models by \citet{thmabe03} as
done for the central values. The gradients of the properties of the
bulge stellar population were derived from the values of age, metallicity, and
$\alpha/$Fe enhancement in the radial range out to \Rbd. The errors on
the gradients were calculated through Monte Carlo simulations taking
into account the errors on the values out to \Rbd\/. The final
gradients of the age, metallicity, and $\alpha/$Fe enhancement and
their corresponding errors are listed in
Tab. \ref{tab:agemetalfa_grad}. The histograms of their number
distribution are plotted in Fig. \ref{fig:histgrad_ama}.

%%%%%%%%%%%%%%%%%%%%%%%%%%%%%%%%%%%%%%%%%%%%%%%%%%%%%%%%%%%%%%%%%%%%%%%%%
%%% TABLE age metallicity alfa/fe gradients  values
\begin{table*}
\caption{Gradients of age, metallicity, and $\alpha/$Fe enhancement of
  the stellar populations of the sample bulges derived from the
  central values and values at the radius \Rbd\ where the
  surface-brightness contributions of the bulge and disc are equal}
\begin{center}
\begin{small}
\begin{tabular}{lrrrr}
\hline
\noalign{\smallskip}
\multicolumn{1}{c}{Galaxy} &
\multicolumn{1}{c}{\Rbd} &
\multicolumn{1}{c}{$\Delta$(Age)} &
\multicolumn{1}{c}{$\Delta$(\ZH)} &
\multicolumn{1}{c}{$\Delta$(\aFe)}\\
\noalign{\smallskip}
\multicolumn{1}{c}{} &
\multicolumn{1}{c}{(arcsec)} &
\multicolumn{1}{c}{[Gyr]} &
\multicolumn{1}{c}{} &
\multicolumn{1}{c}{} \\
\noalign{\smallskip}
\multicolumn{1}{c}{(1)} &
\multicolumn{1}{c}{(2)} &
\multicolumn{1}{c}{(3)} &
\multicolumn{1}{c}{(4)} &
\multicolumn{1}{c}{(5)}  \\
\noalign{\smallskip}
\hline
\noalign{\smallskip}  
ESO-LV~1860550  & $16.2$ & $ 2.01\pm2.14$ & $-0.05\pm0.05$ & $-0.06\pm0.12$ \\
ESO-LV~2060140  & $ 2.2$ & $-0.46\pm0.26$ & $ 0.10\pm0.06$ & $-0.03\pm0.15$ \\
ESO-LV~2340130  & $14.2$ & $-0.33\pm0.85$ & $-0.18\pm0.13$ & $-0.11\pm0.19$ \\
ESO-LV~4000370  & $ 1.5$ & $ 1.19\pm1.96$ & $-0.28\pm0.12$ & $ 0.13\pm0.15$ \\
ESO-LV~4880490  & $23.3$ & $-0.47\pm1.50$ & $ 0.07\pm0.16$ & $ 0.07\pm0.25$ \\
ESO-LV~5340200  & $ 9.8$ & $ 0.03\pm0.18$ & $ 0.15\pm0.06$ & $-0.02\pm0.08$ \\
PGC~26148       & $ 1.5$ & $ 0.78\pm0.39$ & $-0.22\pm0.07$ & $-0.45\pm0.10$ \\
PGC~37759       & $ 1.1$ & $ 0.57\pm1.25$ & $-0.13\pm0.07$ & $-0.07\pm0.17$ \\
\noalign{\smallskip}
\hline
\noalign{\medskip}
\end{tabular}
\end{small}
%\begin{minipage}{8cm}
%NOTE -- 
%Col.(1):Name; Col.(2): metallicity
%Col.(4):
%\end{minipage}
\label{tab:agemetalfa_grad}
\end{center}
\end{table*}

%%%%%%%%%%%%%%%%%%%%%%%%%%%%%%%%%%%%%%%%%%%%%%%%%%%%%%%%%%%%

%%%%%%%%%%%%%%%%%%%%%%%%%%%%%%%%%%%%%%%%%%%%%%%%%%%%%%%%%%%%
% Figure histogramma valori gradienti eta metallicita alfa enhancement

%%%%%%%%%%%%%%%%%%%%%%%%%%%%%%%%%%%%%%%%%%%%%%%%%%%%%%%%%%%%

%%%%%%%%%%%%%%%%%%%%%%%%%%%%%%%%%%%%%%%%%%%%%%%%%%%%%%%%%%%%
% --- Figure gradienti met-grad(met) alfafe-grad(alfafe) ------------------------------------------------------

\begin{figure*}%[hp!]
\centering
\includegraphics[angle=90,width=0.95\textwidth]{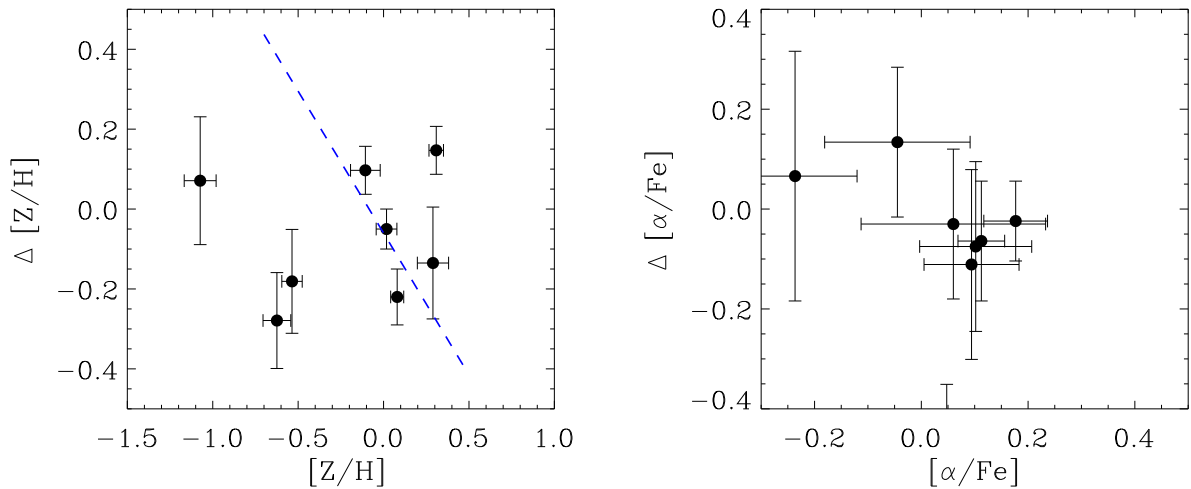}\\
\caption[]{Gradient and central value of metallicity (left-hand panel)
  and $\alpha$/Fe enhancement (right-hand panel) for the sample
  bulges. In the left-hand panel the dashed blue line represents the
  linear regression obtained for bulges of HSB galaxies by
  \citet{moreetal08}}
\label{fig:grad_central_met_alfa}
\end{figure*}

%%%%%%%%%%%%%%%%%%%%%%%%%%%%%%%%%%%%%%%%%%%%%%%%%%%%%%%%%%%%%%%%%%%%%%%%%

All the sample bulges show no age gradient within the errorbars,
except for ESO-LV~2060140 and PGC~26148 which are characterized by a
shallow gradient. At face values the largest age gradients are
measured for the bulges of ESO-LV~1860550, ESO-LV~400370,
($\Delta$(age)$\,=\,1-2$ Gyr, Fig. \ref{fig:histgrad_ama}, left-hand panel), which are
nevertheless consistent with $\Delta$(age)$\,=\,0$ due to their large
uncertainties (Tab. \ref{tab:agemetalfa_grad}).

In spite of the peak at $\Delta($\ZH$)\,=\,-0.15$, the median of the
number distribution of the metallicity gradients is consistent with
$\Delta($\ZH$)\,=\,0$ (Fig. \ref{fig:histgrad_ama}, middle panel). Only
ESO-LV~2340130, ESO-LV~4000370 and PGC~26148 host a bulge with a
stellar population characterized by a metallicity gradient which is
significantly negative (Tab. \ref{tab:agemetalfa_grad}).
Also the number distribution of the gradients of $\alpha/$Fe
enhancements has a median $\Delta($\aFe$)\,=\,0$
(Fig. \ref{fig:histgrad_ama}, right-hand panel).
In all the observed distributions, most of the deviations from the
median values can be explained as due only to the errors in the
estimates of the gradients (Tab. \ref{tab:agemetalfa_grad}).

The absence of significant age and $\alpha/$Fe gradients is in
agreement with the earlier findings for early-type galaxies
\citep{mehletal03, sancetal06s, spoletal10} and bulges of unbarred
\citep{jabletal07, moreetal08} and barred \citep{sancetal11,adri12} HSB galaxies. On the other hand, negative gradients of
metallicity are observed in the radial profiles of many early-type
galaxies \citep{procetal02, mehletal03, sancetal06s, rawletal10} and
bulges of HSB galaxies \citep{jabletal07, moreetal08}.

In the models the presence of a negative metallicity gradient predicts
a formation scenario of the bulges in LSB galaxies via dissipative
collapse \citep{eglbsa62,lars74,aryo87} when a strong interplay
between the star formation time-scale and gas flows is taken into
account to explain of the absence of any $\alpha/$Fe gradient
\citep{pipietal08}. But, pure dissipative collapse is not able to
explain formation of all the sample bulges and other phenomena, like
mergers or acquisition events, need to be invoked to account for the
formation of those bulges which do not show any metallicity gradient
\citep{besh99, cobari99}.

The metallicity gradients are plotted as a function of the metallicity
in the galaxy centre in the left-hand panel of Fig.
\ref{fig:grad_central_met_alfa}. Only the bulges of LSB galaxies with
higher metallicity ($-0.1\,<\,$\ZH$ <\,0.4$) are consistent with the
correlation found by \citet{moreetal08} and \citet{rawletal10} for the
bulges in HSB galaxies and early-type galaxies, respectively. This is
not the case of the few remaining bulges with a very low central
metallicity value ($-1.1\,< $\ZH$ <\,-0.5$). If confirmed with more firm
statistics, such a result would favor the importance of dissipative
collapse in the assembly of bulges \citep{aryo87,pipietal10}.

No correlation has been found between the central value and gradient
of $\alpha$/Fe enhancement the right-hand panel of
Fig. \ref{fig:grad_central_met_alfa}. This is in agreement with the
earlier findings for early-type galaxies \citet{redaetal07} and spiral
galaxies \citep{ moreetal08} and it is expected due to the absence of
gradients in the $\alpha/$Fe radial profiles of the sample galaxies.

\section{Conclusions}
\label{sec:conclusions}

The properties of the stellar population of the bulges of a sample of 8
 spiral galaxies with LSB discs were investigated to constrain the dominant
mechanism at the epoch of their assembly. 

\begin{itemize} 

\item The central values of the velocity dispersion $\sigma$ and \Mgb,
  \Mgd, \Hb, \Fe, and \MgFe\/ line-strength indices were derived from
  the available spectra for all the sample galaxies. The correlations
  between \Mgd , \Fe , \Hb , and $\sigma$ were found to be consistent
  (and in the case of \Mgd\/ remarkably similar) to the relations
  obtained for early-type galaxies and bulges of HSB galaxies
  \citep{idiaetal96, prugetal01, procetal02, gandetal07, moreetal08}.

\item The mean age, total metallicity, and total $\alpha/$Fe
  enhancement of the stellar population in the bulge-dominated region of the
  sample galaxies were derived by using the stellar population models
  by \citet{thmabe03}. The studied bulges are characterized by a very
  young stellar population, with a distribution of ages peaked at the
  value of 1.5 Gyr. They are characterized by ongoing star formation,
  confirming previous studies on a few LSB galaxies \citep{bergetal03}. 
  The metallicity of the sample bulges spans a large range of values
  from high (\ZH$\,=\,0.30$ dex) to sub-solar metallicity
  (\ZH$\,=\,-1.0$ dex).  Most of them display solar $\alpha/$Fe
  enhancements. These properties resemble closely the properties of
  the bulges hosted in HSB galaxies and suggest a formation through a
  dissipative collapse with a short star-formation time-scale
  \citep{thometal05}.
  For the galaxies with sub-solar values of the $\alpha/$Fe
  enhancement, other mechanism of bulge formation as redistribution of
  disc material due to the presence of a bar or
  environmental effects \citep{korken04,mooretal06} need
  to be considered.

\item We do not find any correlation between the age, metallicity, and
  $\alpha/$Fe enhancement and morphological type. Although we
  lack of S0 galaxies because our galaxies ranges from Sa to Sm, we
  can exclude a strong interplay between the bulge and disc
  components.
  There is a correlation between the velocity dispersion, age,
  metallicity, and $\alpha/$Fe enhancement. In agreement with the
  bulges of the LSB discs the most massive bulges of our sample
  galaxies, more metal rich, and characterized by a shorter
  star-formation time-scale.

\item The bulge of the barred galaxy ESO-LV~4880490
% can be classified as a pseudobulge, according to its morphological,
%  photometrical, and kinematical properties. 
  has an intermediate age (3 Gyr), low metallicity, (\ZH$=-1.07$ dex)
  and sub-solar $\alpha/$Fe ratio (\aFe$=-0.24$ dex). These properties
  are consistent with a slow buildup within a scenario of secular
  evolution driven by the bar.

\item Most of the sample galaxies show no gradient in age and
  \aFe\/ radial profiles. This is in agreement with the earlier
  findings by \citet{jabletal07, moreetal08} for the bulges of HSB
  galaxies.
  The presence of negative gradient in the metallicity radial profile
  in some of the sample bulges suggest a formation via dissipative
  collapse \citep{eglbsa62,lars74}. In this framework, a strong interplay between the
  star-formation time-scale and gas flows is needed to explain the
  coexistence of the metallicity gradient with the absence of
  $\alpha/$Fe gradient \citep{pipietal08}, as observed in our data. However, in
  most of our bulges no metallicity gradient is measured. This
  suggests that a pure dissipative collapse is not able to explain
  formation of the bulges of all the LSB galaxies and that other
  phenomena, like mergers or acquisition events, need to be invoked
  \citep{besh99, cobari99}. This picture is also supported by lack of
  correlation between the central value and gradient of metallicity in
  bulges with very low metallicity.

\end{itemize} 

In this work, we highlighted that the bulges hosted by LSB galaxies
share many structural and chemical properties with the bulges of HSB
galaxies. Such a similarity suggests that they possibly had common
formation scenarios and evolution histories. Our findings are in
agreement with and extend previous results inferred by
\citet{mcghetal95} and \citet{beijetal99}, who compared the
photometric properties of the bulges of LSB and HSB galaxies, and
\citet{coccetal08} who performed a detailed analysis of the
kinematical and mass-distribution properties of the bulge of the LSB
galaxy ESO~323-G064.
The fact that bulges hosted in galaxies with very different discs
are remarkably similar rules out a relevant interplay between the
bulge and disc components and give further support to earlier
findings \citep{thda06, moreetal08}. To be definitely confirmed, this
prediction requires the detailed comparison between the properties of
the stellar populations of both bulges and discs in LSB galaxies. 

\section*{Acknowledgments}
This work was supported by Padua University through the grants
CPDA089220/08, 60A02-5934/09, and 60A02-1283/10 and by Italian Space
Agency through the grant ASI-INAF I/009/10/0.  L.M. acknowledges
financial support from Padua University grant CPS0204. JMA is
partially funded by the Spanish MICINN under the Consolider-Ingenio
2010 Program grant CSD2006-00070 and by the Spanish MICINN (grants
AYA2007-67965-C03-01 and AYA2010-21887-C04-04). M.C acknowledges
financial support from Padua University grants CPDR095001/09 and
CPDR115539/11. L.C. has received funding from the European Community's
SeventhFramework Programme (/FP7/2007-2013/) under grant agreement No
229517.

\bibliographystyle{mn2e} \bibliography{bibl}

\appendix
%%%%%%%%%%%%%%%%%%%%%%%%%%%%%%%%%%%%%%%%%%%%%%%%%%%%%%%%%%%%%%%%%%%%%%%%%
%%% KINEMATICAL and indices table

\clearpage
\section{Stellar kinematics and line strength indices.}
The stellar kinematics of PGC~26148 and PGC~37759 and the
line-strength indices for all the sample galaxies are given in Table
A1 and A2 respectively. The stellar kinematics of the remaining sample
galaxies is given in Paper I.

\begin{table*}
\caption{Stellar kinematics of the sample galaxies.}
\begin{tabular}{rrrrr}
\hline
\noalign{\smallskip}
\multicolumn{1}{r}{r~~~~~~~} &
\multicolumn{1}{c}{$v$}&
\multicolumn{1}{c}{$\sigma$} &
\multicolumn{1}{c}{$h_3$} &
\multicolumn{1}{c}{$h_4$}  \\
\multicolumn{1}{r}{[arcsec]} &
\multicolumn{1}{c}{(\AA)} &
\multicolumn{1}{c}{(mag)} &
\multicolumn{1}{c}{(\AA)} &
\multicolumn{1}{c}{(\AA)} \\
\multicolumn{1}{r}{(1)~~~~~} &
\multicolumn{1}{c}{(2)} &
\multicolumn{1}{c}{(3)} &
\multicolumn{1}{c}{(4)} &
\multicolumn{1}{c}{(5)} \\
\noalign{\smallskip}
\hline
\noalign{\smallskip}
{\bf PGC~26148 - MJ} \\
\noalign{\smallskip}
\hline
\noalign{\smallskip}
$-7.43 $ & $    -3.1  \pm     2.4 $ & $     42.8 \pm      2.7 $ & $   -0.004 \pm    0.036  $ & $   0.026 \pm    0.041$ \\
$-4.70 $ & $    14.5  \pm     2.4 $ & $     70.9 \pm      1.9 $ & $   -0.069 \pm    0.018  $ & $  -0.006 \pm    0.022$ \\
$-2.40 $ & $    13.2  \pm     2.6 $ & $     94.7 \pm      2.2 $ & $   -0.042 \pm    0.016  $ & $  -0.014 \pm    0.018$ \\
$-1.19 $ & $     0.1  \pm     2.0 $ & $    103.8 \pm      2.4 $ & $    0.010 \pm    0.015  $ & $   0.016 \pm    0.015$ \\
\end{tabular}
\end{table*}
%%%%%%%%%%%%%%%%%%%%%%%%%%%%%%%%%%%%%%%%%%%%%%%%%%%%%%%%%%%%%%%%%%%%%%%%%%%%%%%%

\begin{table*}
\caption{Line-strength indices of the sample galaxies.  }
\begin{tabular}{rrrrrrrr}
\hline
\noalign{\smallskip}
\multicolumn{1}{r}{r~~~~~~~} &
\multicolumn{1}{c}{\Hb}&
\multicolumn{1}{c}{\Mgd} &
\multicolumn{1}{c}{\Mgb} &
\multicolumn{1}{c}{{\rm Fe5270}} &
\multicolumn{1}{c}{{\rm Fe5335}}  \\
\multicolumn{1}{r}{[arcsec] } &
\multicolumn{1}{c}{(\AA)} &
\multicolumn{1}{c}{(mag)} &
\multicolumn{1}{c}{(\AA)} &
\multicolumn{1}{c}{(\AA)} &
\multicolumn{1}{c}{(\AA)} \\
\multicolumn{1}{r}{(1)~~~~~} &
\multicolumn{1}{c}{(2)} &
\multicolumn{1}{c}{(3)} &
\multicolumn{1}{c}{(4)} &
\multicolumn{1}{c}{(5)} &
\multicolumn{1}{c}{(6)} \\
\noalign{\smallskip}
\hline
\noalign{\smallskip}
{\bf ESO-LV~1860550 - MJ+30$^\circ$} \\
\noalign{\smallskip}
\hline
\noalign{\smallskip}
$-10.88 $ & $ 3.347  \pm   0.134  $ & $    0.131 \pm    0.004   $ & $   2.501  \pm   0.126  $ & $    2.164  \pm   0.149   $ & $   1.707 \pm    0.139 $ \\ 
$-7.70  $ & $ 2.666  \pm   0.116  $ & $    0.153 \pm    0.003   $ & $   2.843  \pm   0.108  $ & $    2.399  \pm   0.127   $ & $   1.889 \pm    0.117 $ \\ 
$-5.58  $ & $ 2.275  \pm   0.105  $ & $    0.177 \pm    0.003   $ & $   3.100  \pm   0.100  $ & $    2.310  \pm   0.118   $ & $   1.958 \pm    0.109 $ \\ 
$-4.09  $ & $ 2.092  \pm   0.093  $ & $    0.185 \pm    0.003   $ & $   3.175  \pm   0.087  $ & $    2.661  \pm   0.101   $ & $   2.253 \pm    0.094 $ \\
\end{tabular}
\end{table*}
\begin{table*}
\contcaption{}
\begin{tabular}{rrrrrrrrr}
\noalign{\smallskip}
\hline
\multicolumn{1}{r}{(1)~~~~} &
\multicolumn{1}{c}{(2)} &
\multicolumn{1}{c}{(3)} &
\multicolumn{1}{c}{(4)} &
\multicolumn{1}{c}{(5)} &
\multicolumn{1}{c}{(6)} \\
\noalign{\smallskip}
\hline
\noalign{\smallskip}
$-13.98 $ & $    3.464  \pm    0.331  $ & $    0.081 \pm    0.010  $ & $    1.977 \pm    0.296   $ & $   1.776 \pm    0.426  $ & $    1.407 \pm   0.286 $ \\ 
$-10.08 $ & $    3.346  \pm    0.215  $ & $    0.093 \pm    0.007  $ & $    1.959 \pm    0.193   $ & $   1.689 \pm    0.282  $ & $    1.358 \pm   0.195 $ \\ 
$-7.45 $ & $    3.189  \pm    0.188  $ & $    0.106 \pm    0.006  $ & $    2.151 \pm    0.172   $ & $   1.808 \pm    0.248  $ & $    1.676 \pm   0.172 $ \\ 
$-5.20 $ & $    3.096  \pm    0.172  $ & $    0.100 \pm    0.005  $ & $    1.934 \pm    0.158   $ & $   1.747 \pm    0.231  $ & $    1.768 \pm   0.158 $ \\ 
$-3.45 $ & $    2.882  \pm    0.169  $ & $    0.106 \pm    0.005  $ & $    1.971 \pm    0.155   $ & $   2.066 \pm    0.224  $ & $    1.941 \pm   0.153 $ \\ 
$-2.20 $ & $    2.928  \pm    0.134  $ & $    0.112 \pm    0.005  $ & $    2.089 \pm    0.131   $ & $   1.997 \pm    0.192  $ & $    1.920 \pm   0.131 $ \\ 
$-1.34 $ & $    3.513  \pm    0.128  $ & $    0.117 \pm    0.004  $ & $    2.005 \pm    0.133   $ & $   2.085 \pm    0.196  $ & $    1.652 \pm   0.131 $ \\
\end{tabular}
\end{table*}
\begin{table*}
\contcaption{}
\begin{tabular}{rrrrrrrrr}
\noalign{\smallskip}
\hline
\multicolumn{1}{r}{(1)~~~~} &
\multicolumn{1}{c}{(2)} &
\multicolumn{1}{c}{(3)} &
\multicolumn{1}{c}{(4)} &
\multicolumn{1}{c}{(5)} &
\multicolumn{1}{c}{(6)} \\
\noalign{\smallskip}
\hline
\noalign{\smallskip}
$ 6.35 $ & $    3.676 \pm    0.167 $ & $    0.101 \pm     0.004  $ & $   1.962 \pm     0.171 $ & $    1.735  \pm    0.186  $ & $   1.461  \pm    0.118 $ \\
$ 8.31 $ & $    3.606 \pm    0.161 $ & $    0.109 \pm     0.004  $ & $   2.205 \pm     0.171 $ & $    1.709  \pm    0.192  $ & $   1.669  \pm    0.126 $ \\
$11.17 $ & $    3.686 \pm    0.136 $ & $    0.092 \pm     0.004  $ & $   1.884 \pm     0.150 $ & $    1.564  \pm    0.173  $ & $   1.498  \pm    0.117 $ \\
$16.35 $ & $    3.601 \pm    0.204 $ & $    0.085 \pm     0.006  $ & $   1.677 \pm     0.223 $ & $    0.910  \pm    0.254  $ & $   1.375  \pm    0.176 $ \\
\noalign{\smallskip}
\hline
\noalign{\smallskip}
{\bf ESO-LV~2340130 - MN+16$^\circ$} \\
\noalign{\smallskip}
\hline
\noalign{\smallskip}
$-7.35 $ & $    3.677 \pm    0.155 $ & $    0.077 \pm    0.004 $ & $    2.092 \pm    0.133 $ & $    1.399  \pm   0.157  $ & $   1.226 \pm    0.144$ \\
$-3.90 $ & $    3.621 \pm    0.129 $ & $    0.096 \pm    0.003 $ & $    1.777 \pm    0.103 $ & $    1.597  \pm   0.120  $ & $   1.377 \pm    0.111$ \\
\end{tabular}
\end{table*}
\begin{table*}
\contcaption{}
\begin{tabular}{rrrrrrrrr}
\noalign{\smallskip}
\hline
\multicolumn{1}{r}{(1)~~~~} &
\multicolumn{1}{c}{(2)} &
\multicolumn{1}{c}{(3)} &
\multicolumn{1}{c}{(4)} &
\multicolumn{1}{c}{(5)} &
\multicolumn{1}{c}{(6)} \\
\noalign{\smallskip}
\hline
\noalign{\smallskip}
$-6.26 $ & $    3.135 \pm    0.320 $ & $    0.068 \pm    0.009 $ & $    1.080  \pm   0.238  $ & $   1.374 \pm    0.288 $ & $    1.693 \pm    0.296$ \\
$-4.50 $ & $    3.638 \pm    0.296 $ & $    0.085 \pm    0.008 $ & $    1.694  \pm   0.216  $ & $   1.964 \pm    0.262 $ & $    1.518 \pm    0.268$ \\
$-3.26 $ & $    2.881 \pm    0.288 $ & $    0.077 \pm    0.008 $ & $    1.878  \pm   0.214  $ & $   1.546 \pm    0.258 $ & $    1.849 \pm    0.258$ \\
$-2.39 $ & $    2.719 \pm    0.275 $ & $    0.083 \pm    0.008 $ & $    1.561  \pm   0.204  $ & $   1.581 \pm    0.246 $ & $    1.632 \pm    0.245$ \\
$-1.77 $ & $    2.963 \pm    0.275 $ & $    0.084 \pm    0.008 $ & $    1.530  \pm   0.207  $ & $   1.693 \pm    0.249 $ & $    1.647 \pm    0.248$ \\
$-1.27 $ & $    2.493 \pm    0.252 $ & $    0.079 \pm    0.007 $ & $    1.508  \pm   0.192  $ & $   2.032 \pm    0.227 $ & $    1.789 \pm    0.229$ \\
$-0.78 $ & $    2.354 \pm    0.246 $ & $    0.098 \pm    0.007 $ & $    1.902  \pm   0.179  $ & $   2.109 \pm    0.221 $ & $    1.664 \pm    0.222$ \\
\end{tabular}
\end{table*}
\begin{table*}
\contcaption{}
\begin{tabular}{rrrrrrrrr}
\noalign{\smallskip}
\hline
\multicolumn{1}{r}{(1)~~~~} &
\multicolumn{1}{c}{(2)} &
\multicolumn{1}{c}{(3)} &
\multicolumn{1}{c}{(4)} &
\multicolumn{1}{c}{(5)} &
\multicolumn{1}{c}{(6)} \\
\noalign{\smallskip}
\hline
\noalign{\smallskip}
$  -1.37  $ & $   2.890 \pm    0.118 $ & $    0.146 \pm    0.004 $ & $    2.902 \pm    0.109 $ & $  2.297 \pm   0.137 $ & $    1.714 \pm    0.122 $ \\ 
$  -1.01  $ & $   2.985 \pm    0.115 $ & $    0.148 \pm    0.004 $ & $    2.493 \pm    0.106 $ & $  2.286 \pm   0.134 $ & $    1.844 \pm    0.124 $ \\ 
$  -0.76  $ & $   2.950 \pm    0.122 $ & $    0.155 \pm    0.004 $ & $    2.652 \pm    0.113 $ & $  2.392 \pm   0.142 $ & $    2.018 \pm    0.133 $ \\ 
$  -0.51  $ & $   2.942 \pm    0.104 $ & $    0.159 \pm    0.004 $ & $    3.009 \pm    0.099 $ & $  2.415 \pm   0.126 $ & $    2.024 \pm    0.117 $ \\ 
$  -0.26  $ & $   2.746 \pm    0.085 $ & $    0.178 \pm    0.003 $ & $    3.327 \pm    0.083 $ & $  2.523 \pm   0.106 $ & $    2.126 \pm    0.099 $ \\ 
$  -0.01  $ & $   2.980 \pm    0.074 $ & $    0.178 \pm    0.003 $ & $    3.162 \pm    0.073 $ & $  2.637 \pm   0.094 $ & $    2.103 \pm    0.090 $ \\ 
$   0.24  $ & $   2.932 \pm    0.071 $ & $    0.183 \pm    0.003 $ & $    2.983 \pm    0.070 $ & $  2.581 \pm   0.090 $ & $    2.084 \pm    0.086 $ \\
\end{tabular}
\end{table*}
\begin{table*}
\contcaption{}
\begin{tabular}{rrrrrrrrr}
\noalign{\smallskip}
\hline
\multicolumn{1}{r}{(1)~~~~} &
\multicolumn{1}{c}{(2)} &
\multicolumn{1}{c}{(3)} &
\multicolumn{1}{c}{(4)} &
\multicolumn{1}{c}{(5)} &
\multicolumn{1}{c}{(6)} \\
\noalign{\smallskip}
\hline
\noalign{\smallskip}
$  -1.22  $ & $   2.693 \pm    0.194 $ & $    0.059 \pm    0.006 $ & $    3.535 \pm    0.214 $ & $    2.738 \pm    0.281 $ & $    2.049 \pm    0.261 $ \\
$  -0.61  $ & $   3.087 \pm    0.161 $ & $    0.049 \pm    0.005 $ & $    3.003 \pm    0.174 $ & $    2.698 \pm    0.228 $ & $    1.987 \pm    0.214 $ \\
$  -0.25  $ & $   3.097 \pm    0.185 $ & $    0.036 \pm    0.006 $ & $    3.021 \pm    0.193 $ & $    3.052 \pm    0.256 $ & $    2.061 \pm    0.243 $ \\
$   0.00  $ & $   2.267 \pm    0.175 $ & $    0.055 \pm    0.005 $ & $    2.932 \pm    0.182 $ & $    2.625 \pm    0.241 $ & $    2.149 \pm    0.231 $ \\
$   0.25  $ & $   2.997 \pm    0.163 $ & $    0.054 \pm    0.005 $ & $    2.862 \pm    0.169 $ & $    2.402 \pm    0.221 $ & $    2.186 \pm    0.221 $ \\
$   0.50  $ & $   3.225 \pm    0.180 $ & $    0.062 \pm    0.006 $ & $    3.072 \pm    0.186 $ & $    2.506 \pm    0.244 $ & $    2.229 \pm    0.248 $ \\
$   0.86  $ & $   2.859 \pm    0.165 $ & $    0.039 \pm    0.005 $ & $    2.734 \pm    0.169 $ & $    1.789 \pm    0.225 $ & $    2.153 \pm    0.232 $ \\
\end{tabular}
\end{table*}
%%%%%%%%%%%%%%%%%%%%%%%%%%%%%%%%%%%%%%%%%%%%%%%%%%%%%%%%%%%%%%%%%%%%%%%%%%%%%%%%

\bsp

\label{lastpage}

\end{document}